\documentclass[11pt]{article}    
\usepackage{lscape,epsfig,graphicx,multirow}
\usepackage[section]{placeins}
\usepackage{makecell}
\usepackage[affil-it]{authblk}
\usepackage{color}
\title{\textbf{Mass production and characterization of 3-inch PMTs for the JUNO experiment}}
\author[1,2]{Chuanya Cao}
\author[1]{Jilei Xu\thanks{xujl@ihep.ac.cn}}
\author[1]{Miao He\thanks{hem@ihep.ac.cn}}
\author[3]{Angel Abusleme}
\author[4]{Mathieu Bongrand}
\author[5,6]{Cl\'{e}ment Bordereau}
\author[7]{Dominique Breton}
\author[7,8]{Anatael Cabrera}
\author[3]{Agustin Campeny}
\author[5]{C\'{e}dric Cerna}
\author[9]{Haoqiang Chen\thanks{Now at China Electronics Technology Instruments Co., Ltd}}
\author[6]{Po-An Chen}
\author[5]{G\'{e}rard Claverie}
\author[10]{Selma Conforti Di Lorenzo}
\author[10]{Christophe De La Taille}
\author[5]{Fr\'{e}d\'{e}ric Druillole}
\author[5]{Am\'{e}lie Fournier}
\author[7,8]{Marco Grassi}
\author[9]{Xiaofei Gu}
\author[3]{Michael Haacke}
\author[7,8]{Yang Han}
\author[5]{Patrick Hellmuth}
\author[1]{Yuekun Heng}
\author[3]{Rafael Herrera}
\author[6]{Yee Hsiung}
\author[6]{Bei-Zhen Hu}
\author[11]{Yongbo Huang}
\author[5]{C\'{e}dric Huss}
\author[3]{Ignacio Jeria}
\author[1]{Xiaoping Jing}
\author[5]{C\'{e}cile Jollet}
\author[4]{Victor Lebrin}
\author[4]{Fr\'{e}d\'{e}ric Lef\`{e}re}
\author[12]{Hongwei Li}
\author[1]{Nan Li\thanks{Now at Huawei Technologies Co., Ltd}}
\author[11]{Hongbang Liu}
\author[11]{Xiwen Liu}
\author[13]{Bayarto Lubsandorzhiev}
\author[13]{Sultim Lubsandorzhiev}
\author[13]{Arslan Lukanov}
\author[7]{Jihane  Maalmi}
\author[5]{Anselmo Meregaglia}
\author[7]{Diana Navas-Nicolas}
\author[14,3]{Juan Pedro Ochoa-Ricoux}
\author[5]{Fr\'{e}d\'{e}ric Perrot}
\author[4]{Rebin Karaparambil Rajan}
\author[5]{Abdel Rebii}
\author[14]{Bed\v{r}ich Roskovec}
\author[8]{Cayetano Santos}
\author[4]{Mariangela Settimo}
\author[13]{Andrey Sidorenkov}
\author[13]{Igor Tkachev}
\author[3]{Giancarlo Troni}
\author[13]{Nikita Ushakov}
\author[4]{Guillaume Van Royen}
\author[4]{Benoit Viaud}
\author[13]{Dmitriy Voronin}
\author[3]{Pablo Walker}
\author[15]{Chung-Hsiang Wang}
\author[1]{Zhimin Wang}
\author[1,2]{Diru Wu}
\author[1,2]{Hangkun Xu}
\author[1]{Meihang Xu}
\author[11]{Chengfeng Yang}
\author[9]{Jie Yang}
\author[4]{Fr\'{e}d\'{e}ric Yermia}
\author[1]{Xuantong Zhang}

\affil[1]{Institute of High Energy Physics, Beijing, China}
\affil[2]{University of Chinese Academy of Sciences, Beijing 100049, China}
\affil[3]{Pontificia Universidad Cat\'{o}lica de Chile, Santiago, Chile}
\affil[4]{SUBATECH, Universit\'{e} de Nantes,  IMT Atlantique, CNRS-IN2P3, Nantes, France}
\affil[5]{Univ. Bordeaux, CNRS, CENBG, UMR 5797, F-33170 Gradignan, France}
\affil[6]{Department of Physics, National Taiwan University, Taipei}
\affil[7]{IJCLab, Universit\'{e} Paris-Saclay, CNRS/IN2P3, 91405 Orsay, France}
\affil[8]{Astro-Particle Physics Laboratory, CNRS/CEA/Paris7/Observatoire de Paris, Paris, France}
\affil[9]{School of Physics and Microelectronics, Zhengzhou University, Zhengzhou, China}
\affil[10]{OMEGA, Ecole Polytechnique-CNRS/IN2P3, Paris, France}
\affil[11]{Guangxi University, Nanning, China}
\affil[12]{Hainan Zhanchuang Photonics Technology Co., Ltd, Chengmai, China}
\affil[13]{Institute for Nuclear Research of the Russian Academy of Sciences, Moscow, Russia}
\affil[14]{Department of Physics and Astronomy, University of California, Irvine, California 92697, USA}
\affil[15]{National United University, Miao-Li}

\date{2021-2-23}%

\begin{document}
\maketitle
\begin{abstract}

  26,000 3-inch photomultiplier tubes (PMTs) have been produced for Jiangmen Underground Neutrino Observatory (JUNO) by the Hainan Zhanchuang Photonics Technology Co., Ltd (HZC) company in China and passed all acceptance tests with only 15 tubes rejected. The mass production began in 2018 and elapsed for about 2 years at a rate of $\sim$1,000~PMTs per month. The characterization of the PMTs was performed in the factory concurrently with production as a joint effort between HZC and JUNO. Fifteen performance parameters were tracked at different sampling rates, and novel working strategies were implemented to improve quality assurance. This constitutes the largest sample of 3-inch PMTs ever produced and studied in detail to date.  

\end{abstract}
\section{Introduction}

   The Jiangmen Underground Neutrino Observatory (JUNO)~\cite{juno-CDR} is a multipurpose neutrino experiment under construction in southern China. Its main detector is located 53 km from two nuclear power plants in a cavern with a 650~m overburden. The primary goal is to measure the neutrino mass ordering with a sensitivity better than 3 standard deviations after 6 years of data taking~\cite{juno-yellowbook}. High transparency liquid scintillator, high coverage (78$\%$) of photomultiplier tubes (PMTs), and low background levels are needed to achieve an energy resolution of 3$\%$/$\sqrt{E({\rm MeV})}$ and an energy calibration error lower than 1$\%$. The high coverage is achieved by closely packing $\sim$18,000 high quantum efficiency 20-inch PMTs (Large PMTs or LPMTs) around the liquid scintillator target sphere. In addition, $\sim$25,600 3-inch PMTs (Small PMTs or SPMTs) will be installed in the gaps between the LPMTs forming a double calorimetry system~\cite{SPMT-hem}. This system will extend JUNO's physics reach by reducing the systematic uncertainties associated to the energy measurement, improving the reconstruction of muons, and the detection of supernova neutrinos. 

   Small PMTs are widely used in large-scale neutrino physics and astrophysics experiments. For example, KM3NeT~\cite{km3} plans to deploy 200,000 small PMTs in the Mediterranean Sea to make a neutrino telescope. Groups of 31 PMTs hosted within transparent spheres will constitute three-dimensional digital optical modules. So far 7,000 3-inch PMTs have been produced and characterized by KM3NeT~\cite{km3-spmt}.  Hyper-Kamiokande~\cite{HyperK} is considering a hybrid configuration with 20$\%$ coverage of large PMTs and 5$\%$ coverage of small PMTs, and expects to improve the vertex reconstruction and event recognition thanks to the better timing of the latter~\cite{HyperK-spmt}. Recently, LHAASO ordered 2,200 small PMTs to be installed in its water Cherenkov detectors to extend the energy measurement range for astrophysics studies.

   The selection of small PMTs for JUNO started in 2016. A few samples of XP72B20~\cite{XP72B20} from Hainan Zhanchuang Photonics Technology Co., Ltd (HZC) and R12199~\cite{R12199} from Hamamatsu Photonics K.K. were tested. Both of them were found to meet the main requirements of quantum efficiency (QE), single photoelectron (PE) resolution, and so on, which were shown in table 1 of reference~\cite{linan}. At the same time, a modified design of the shape of the glass bulb was designed and produced by HZC with respect to XP72B20 for better timing, leading to a new model, XP72B22. An international bidding was organized in May 2017, and HZC was chosen to be the supplier of all 26,000 XP72B22 PMTs including 400 spares. The mass production of the PMTs started in January 2018, and finished in December 2019, with a production speed of $\sim$1,000 pieces per month. In this paper, we introduce the new features of HZC XP72B22 and its mass production in Sec.~\ref{sec:r_d}. The performance study of the PMT test facilities at HZC is reported in Sec.~\ref{sec:test_station}. The onsite acceptance tests and the quality assurance process followed by JUNO, together with the measured parameters of all 26,000 PMTs are shown in Sec.~\ref{sec:acc_test}.

\section{R\&D of HZC XP72B22 and mass production}
\label{sec:r_d}

\begin{figure}[!hbt]
    \centering  
    \includegraphics[width=0.5\textwidth]{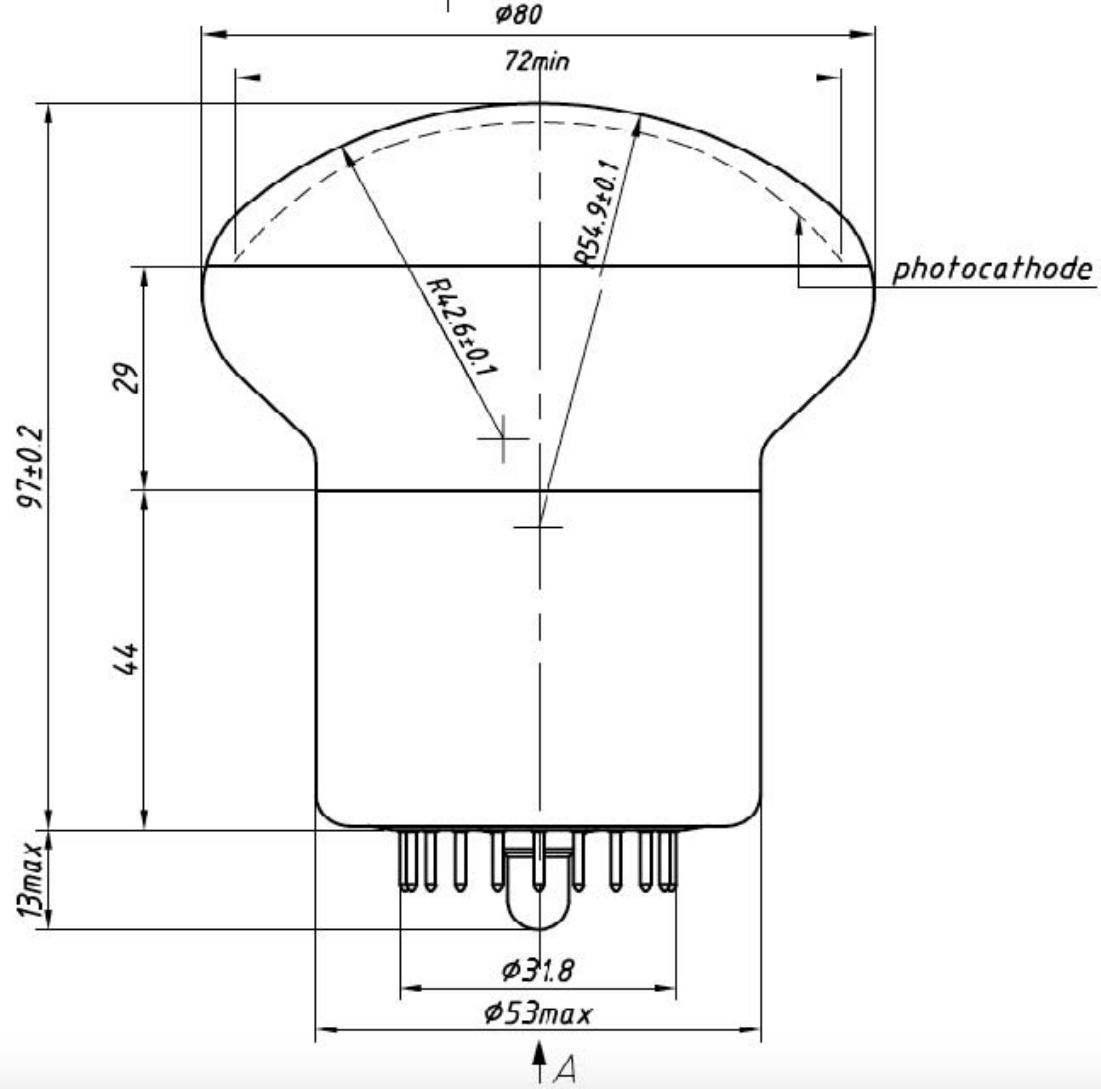}
      \includegraphics[width=0.4\textwidth]{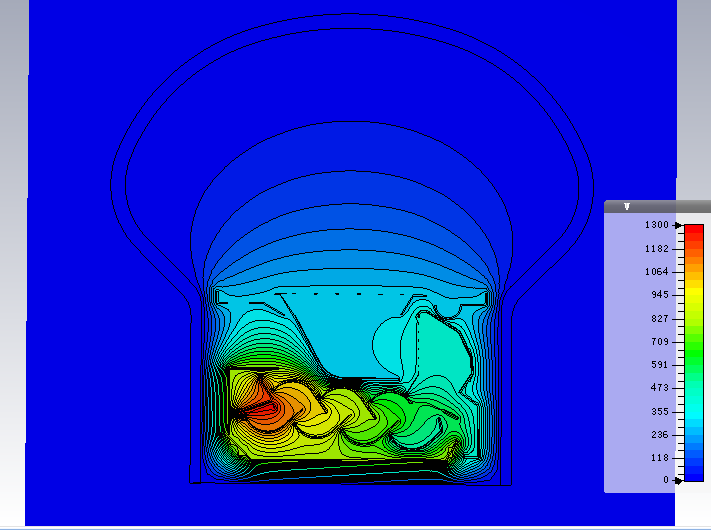}
      \caption{Left: Engineering drawing of PMT XP72B22. Right: Typical electronic field simulation. The dimensions are given in millimeters (left) and the potential in Volts (right).}
   \label{fig:XP72B22}
  \end{figure}

   XP72B20 was originally designed for KM3NeT with the curvature of the photocathode was determined to be 52.4~mm~\cite{XIOPM}. The shape of the glass bulb was further optimized for both collection efficiency and transit time spread (TTS) of photoelectrons (PEs) with simulation study in 2017 at the Xi'an Institute of Optics and Precision Mechanics of the Chinese Academy of Sciences at the request of JUNO. At a given voltage of 265~V which was calculated from gain 3 $\times$ 10$^{6}$ between the photocathode and the first dynode, the electric field distribution was simulated, and the maximum difference of the transit time of PEs emitted at 6 positions with the polar angle from 0$^{\circ}$ to 50$^{\circ}$ was found to be 1.4~ns. A new glass bulb was then designed with a combination of two curvatures: 54.9~mm and 42.6~mm, as shown in Fig.~\ref{fig:XP72B22}. The maximum transit time difference was reduced to 0.5~ns.

   The simulation also indicated that the collection of the multiplied PEs between the first and the second dynode played a significant role in reducing the TTS. The resistor ratio (high voltage ratio) of the first 3 dynodes was originally set to 3:1:1 in an early study of JUNO~\cite{spmt-base}. In order to improve the TTS, a dedicated study was done with different resistor ratios. A ratio of 3:2:1 was finally selected, which gave a 25\% improvement of the TTS, from 5.0~ns to 3.7~ns in terms of full width at half maximum (FWHM) for single PEs. Although the ratio 3:3:1 gave a slightly better TTS, an additional $\sim$50~V (4$\%$) would be required to compensate for the decrease of the gain and the single PE resolution was found to be reduced relatively 5\%.

   As a low-background experiment, the radioactivity of each detector component of JUNO has to be carefully controlled. The requirement on the radioactivity of the glass bulb for the small PMT in JUNO is 400~ppb (4.94~Bq/kg), 400~ppb (1.63~Bq/kg) and 200~ppb (52.47~Bq/kg), for $^{238}$U, $^{232}$Th and $^{40}$K, respectively, based on an investigation of the glass manufacture~\cite{pmt-bkg} and the simulation of the background event rate in the detector~\cite{juno-yellowbook}. The major composition of the glass bulb is quartz sand and 3 different sand samples were obtained from the market and measured by a High Purity Germanium detector. The results are shown in Table~\ref{tab:rawmaterial}. The normal sand has much higher $^{232}$Th than the requirement. $^{238}$U and $^{232}$Th were reduced by a factor of 3 and 20 after acid pickling, resulting in a small cost increase. The high-purity sand yielded another factor of 3 reduction on $^{238}$U and $^{232}$Th, while $^{40}$K was found to be increased significantly probably due to the contamination in the purification procedure. Taking into account the radioactivity and the price, the pickled quartz sand was chosen for the PMT glass bulb production for JUNO. The long-term monitoring of the glass bulb radioactivity will be introduced in Sec.~\ref{sec:radioactivity}.

   Since the JUNO central detector will be immersed in water, radon emanated from materials inside and out of the detector and dissolved in water will constitute another significant source of radioactive background. The required upper limit of radon radioactivity in water is 200~mBq/m$^3$. To evaluate the radon contribution, 29 SPMT glass bulbs were placed into a 700~L large chamber in stainless steel filled with nitrogen to accumulate radon till secular equilibrium was reached. Part of the gas was then pumped into an electrostatic radon detector to measure the alpha particles emitted by radon daughters, especially $^{214}$Po. An introduction to this facility can be found in Refs.~\cite{radon1, radon2}. This measurement gave an emanation rate of $<$ 350 atoms of $^{222}$Rn/day/m$^2$, corresponding to a total contribution from the 25,600 SPMTs of $<$ 0.1~mBq/m$^3$ in the JUNO water pool, which is negligible compared to the requirement.

  \begin{table}
  \caption{Different raw meterial radioactivities and glass bulb radioactivity requirements}
  \vspace{0.2cm}	
  \centering
	\begin{tabular}{cccc}
        \hline
		Raw Material (Bq/kg) & $^{238}$U & $^{232}$Th & $^{40}$K \\
		\hline
        Normal quartz sand & 2.95 $\pm$ 0.09 & 4.07 $\pm$ 0.12 & 8.37 $\pm$ 0.53 \\
		Pickled quartz sand & 1.03 $\pm$ 0.04 & 0.18 $\pm$ 0.02 & 9.35 $\pm$ 0.58 \\
		High-purity sand & 0.29 $\pm$ 0.03 & 0.06 $\pm$ 0.02 & 66.47 $\pm$ 3.59 \\
	    \hline
		Requirement & $<$~4.94 & $<$~1.63 & $<$~52.47 \\
        \hline
	\end{tabular}\label{tab:rawmaterial}
  \end{table}

   The production line of HZC was imported from PHOTONIS France in 2011 with a full production capacity of 250,000 tubes per year. The high degree of automation in both the production line and the performance testing largely ensures the stability of the product quality and reduces the need for human labor and required skills. The quality management system is based on ISO 9001:2005 standards. A dedicated production team was organized and quality control strategies were applied for JUNO. For example, 6 additional steps were implemented for the component inspection. Weekly meetings were organized to analyze product quality issues. In 2017, a pilot production of several hundreds of qualified PMTs was reviewed by JUNO. The quality of these tubes was satisfying and thus the mass production was approved to start at the beginning of 2018. There was no major issue in the entire production period of two years, and the PMTs were supplied to JUNO continuously every three months. The ratio of PMTs that passed the outgoing quality control before delivering to JUNO, defined as the good products yield, was below 50\% in 2017, then increased to 77.5\% in 2018 and 87.8\% in 2019. The average yield was 80.5\%, with the two major sources of disqualification by HZC being low gain and high dark count rate. A further acceptance test by JUNO was done based on the good PMTs, which will be introduced in Sec.~\ref{sec:acc_test}.

\section{Performance study of PMT test stations at HZC}
\label{sec:test_station}

   A waterproof seal will be applied to all 26,000 PMTs together with the HV divider and the cable by HZC. Therefore, an acceptance test by JUNO to ensure the quality of the PMTs was necessary before the sealing. Considering the large number of PMTs, as well as the fact that each of them has 15 parameters (table \ref{tab:criteria}) to be characterized, and in order to reduce the cost, manpower, and risks associated with PMT transportation back and forth, JUNO adopted an onsite sampling test strategy by sending a team to HZC roughly every three months during the production but using the test facilities and the manpower of HZC. This strategy also allowed to inspect the PMTs' performance at an early stage, ensuring good quality control of the production.

   As part of the incoming material inspection, the diameters of the glass bulbs were first measured to ensure they fell into the (78, 82)~mm range. The produced PMTs were measured in four main test stations, which were built or improved before the mass production started, and their performance was reviewed and monitored through the production period.
   \begin{enumerate}
     \item Static station: testing quantum efficiency (QE) and high voltage (HV) at a nominal gain ($3\times10^6$).
     \item Single photoelectron (SPE) station: testing SPE resolution, peak to valley (PV) ratio, dark count rate (DCR).
     \item Transit time spread (TTS) station: testing TTS, pre-pulse, and after-pulse.
     \item Scanning station: testing QE non-uniformity and the effective diameter of the cathode.
   \end{enumerate}

   The first two stations were used by HZC as a standard procedure to test the basic parameters (QE, HV, SPE resolution, PV ratio, DCR) for all PMTs. Only tubes that were qualified during this procedure were given over to JUNO for further testing. All four stations were used by JUNO for the sampling tests.

  \subsection{Static station}

\begin{figure}[!hbt]
     \centering  
     \includegraphics[width=0.5\textwidth]{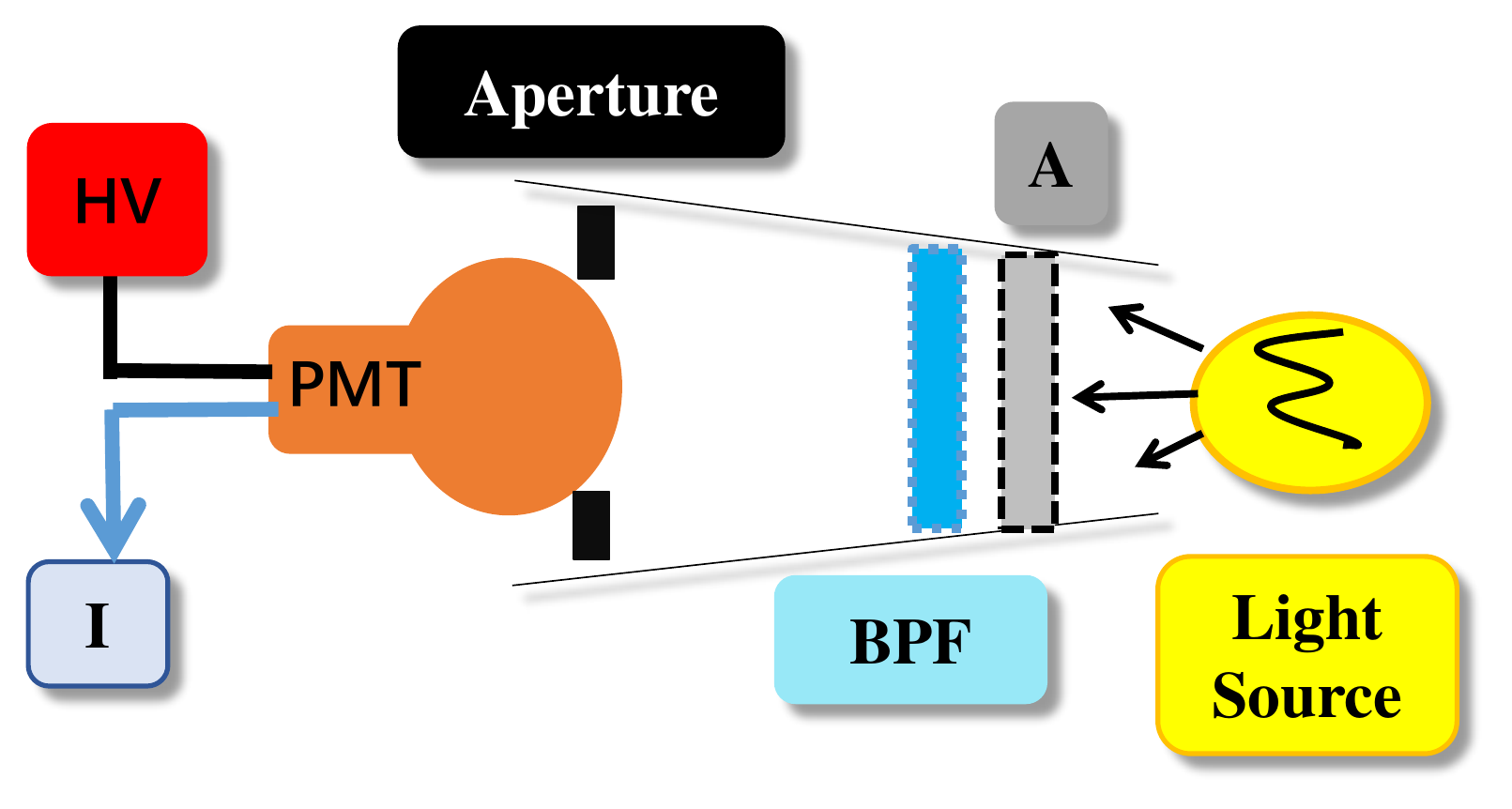}
     \caption{Diagram of the static station to measure QE and HV. The system was in the darkroom. A light spot with 400~nm wavelength and a diameter of 70~mm was provided by a quartz tungsten lamp passed through a band pass filter (BPF) and an aperture. An optical attenuator (A) was added between BPF and light source when measured the anode current.}
     \label{fig:staticST}
   \end{figure}

   The static station (Fig. \ref{fig:staticST}) was used to measure the quantum efficiency (QE) and the high voltage (HV) at nominal gain ($3\times10^6$). Experimentally, QE is defined as the ratio between the photoelectrons produced by photocathode and then collected by the first dynode and the photons emitting into photocathode. However, it is hard to measure the absolute incident photons precisely, so we used a standard PMT to be the reference. For the QE measurement, the light from a quartz tungsten lamp passed through a 400~nm bandpass filter (BPF) and directly hit the cathode with an aperture diameter of 70~mm. The first-dynode current $I_{\rm{k}}$ was read out and compared with the current of a reference PMT $I_{\rm{kc}}$ whose QE$_{\rm{c}}$ was calibrated by a 10~mm $\times$ 20~mm reference photodiode S2744~\cite{PDreference} with the method of Ref.~\cite{QEcalib} with the relative uncertainty of reference PMT QE was estimated about 0.5\%. The QE of the measured PMT was obtained from equation
   
   \begin{equation}\label{eq:QE}
    \rm{QE} = \frac{I_{\rm{k}}}{I_{\rm{kc}}} \rm{QE}_{\rm{c}},
   \end{equation}

   For the HV measurement at such a high gain, an optical attenuator (with attenuation factor A) was added to reduce the anode current $I_{\rm{a}}$ into the range of the ampere meter, and the gain ($G$) was extracted as
   \begin{equation}\label{eq:HV}
    G = \frac{I_{\rm{a}}}{I_{\rm{k}}} A,
   \end{equation}
   where $I_{\rm{k}}$ was measured without attenuation. At nominal gain $G^{\rm nom}$, the corresponding nominal anode current $I^{\rm nom}_{\rm a}$ was calculated using Eq.~(\ref{eq:HV}) and the HV was tuned till $I_{\rm a}$ was close to $I^{\rm nom}_{\rm a}$.
   
   There were three light filters at HZC with wavelengths of 320~nm, 400~nm, and 550~nm, while the QE requirement by JUNO was defined at 420~nm. Therefore, the QE of five XP72B22 PMTs was scanned from 300~nm to 700~nm by JUNO~\cite{linan}, and the average QE at 420~nm was found to be 6.8\% lower than that at 400~nm. A correction factor 0.932 was thus applied to HZC's result at 400~nm and delivered to JUNO. The other two filters were used for the spectral response measurement.

\begin{figure}[!hbt]
  \centering  
  \includegraphics[width=0.45\textwidth]{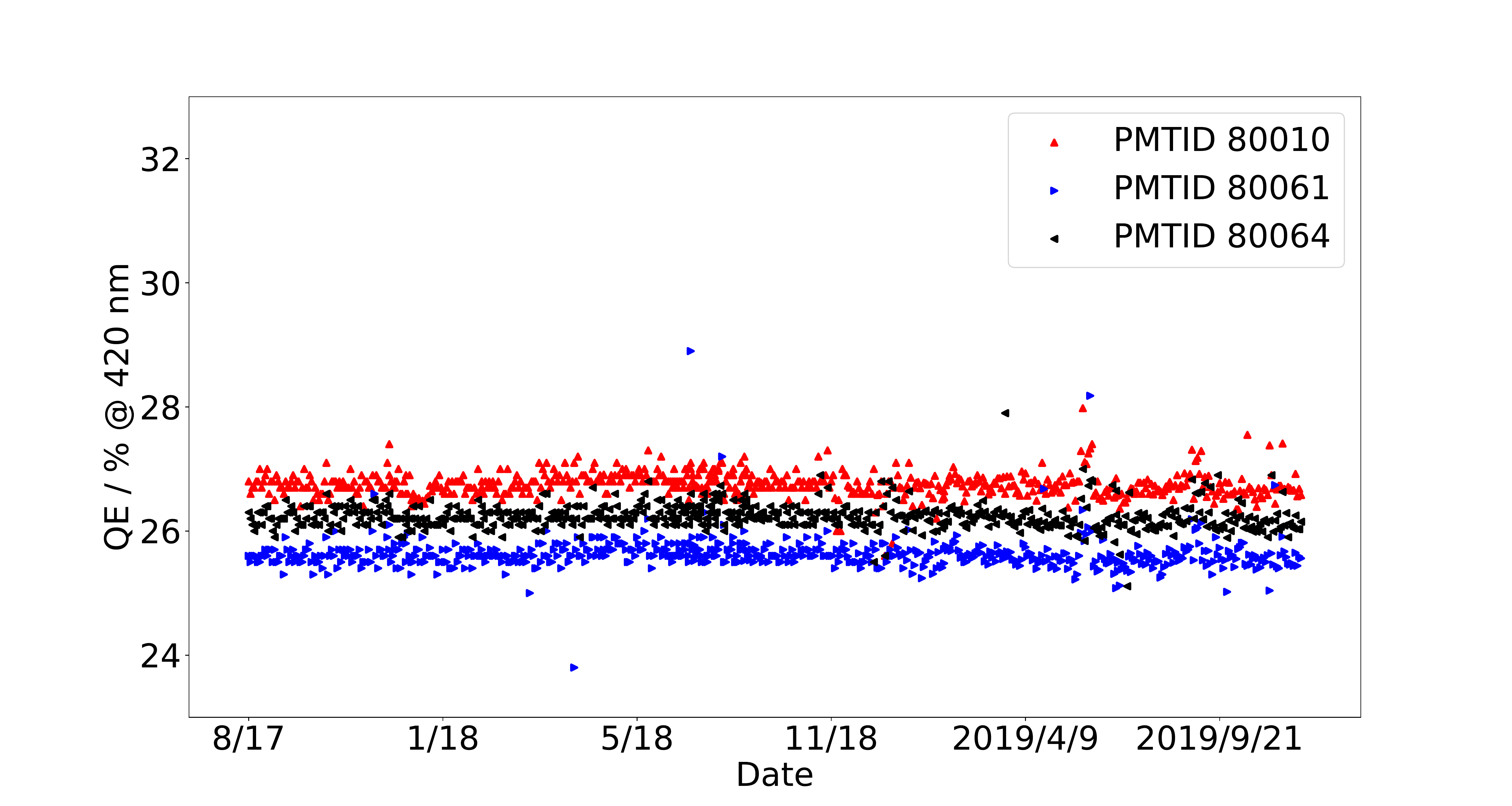}
    \includegraphics[width=0.45\textwidth]{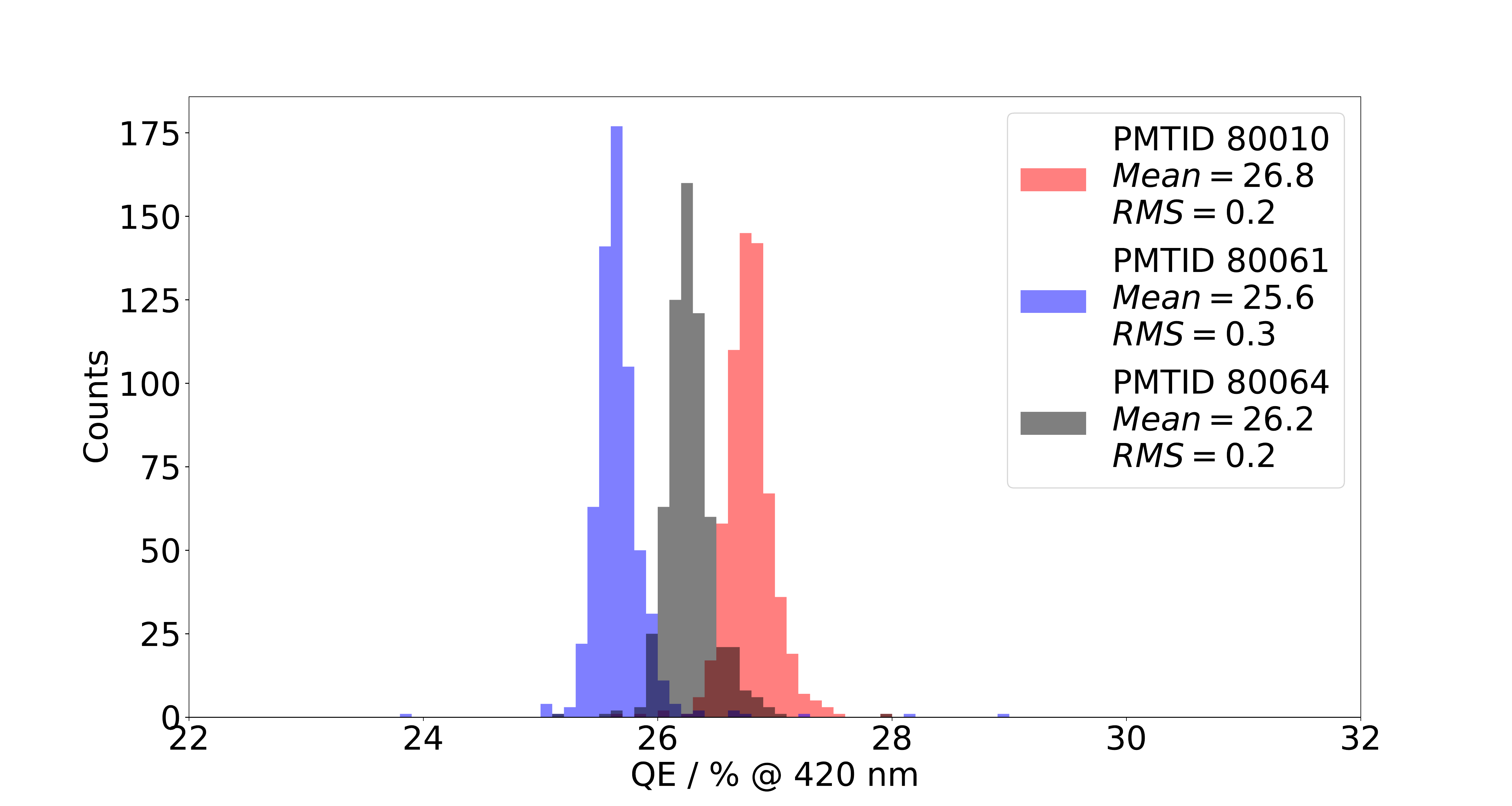}
  \caption{Left: QE monitoring of three PMTs as a function of time. Right: QE distribution for each of the three monitor PMTs.}
  \label{fig:QE-monitor}
\end{figure}

   Three XP72B22 PMTs were measured every day to monitor the working stability of the station during the whole production. As shown in Fig.~\ref{fig:QE-monitor} (left), the QE measurements were stable over the full production period. A few exceptional data points were attributed to the accidental measurement error for a single monitor PMT. The cumulative statistics of QE over the production period is shown in Fig.~\ref{fig:QE-monitor} (right), and their average fluctuation of 0.2\%, corresponding to a relative uncertainty 0.8\%. 

\begin{figure}[!hbt]
  \centering  
  \includegraphics[width=0.45\textwidth]{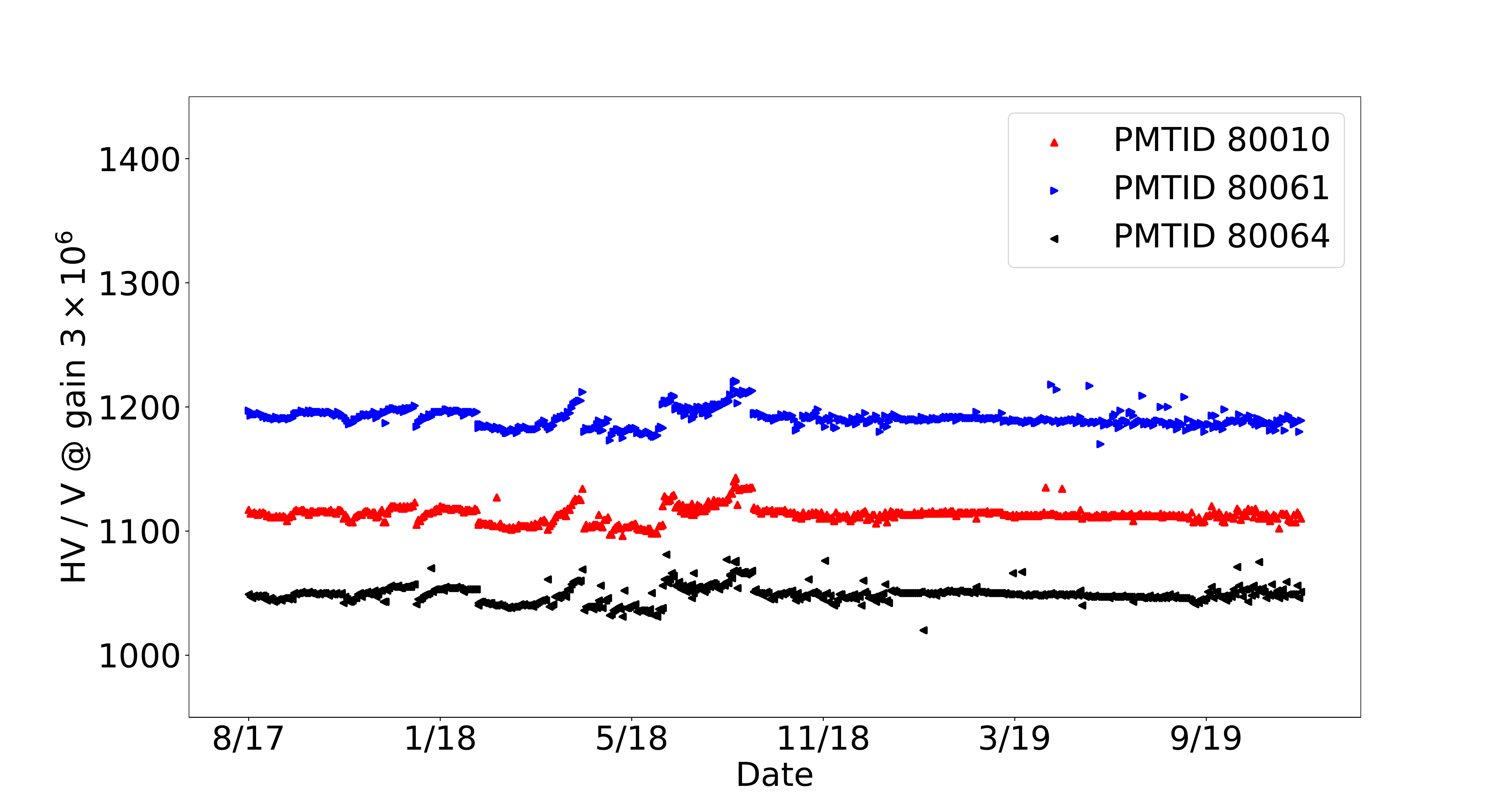}
    \includegraphics[width=0.45\textwidth]{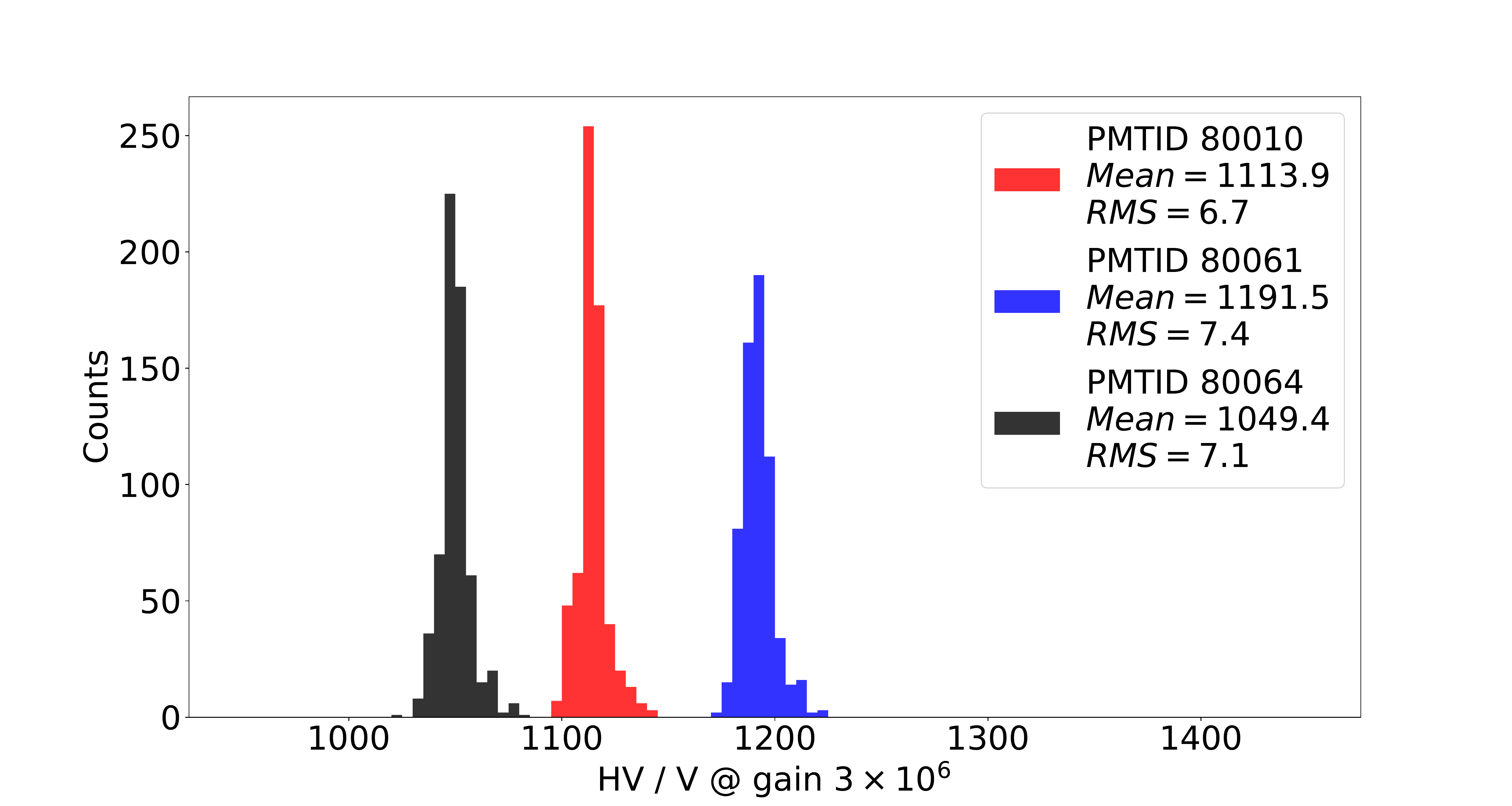}
    \caption{Left: HV monitoring of three PMTs as a function of time. Right: HV distribution for each of the three monitor PMTs.}
  \label{fig:HV-monitor}
\end{figure}

    The HV monitoring data of the same three PMTs are shown in Fig.~\ref{fig:HV-monitor}. There were large fluctuations up to $\pm 20$~V before August 2018. An investigation of the test station suggested some interference between the power supply and the signal readout since they were in the same crate. The power supply was then moved out and the grounding of the readout electronics was improved. As a result, fluctuations were reduced by a factor of three. The three monitor PMTs give similar results, and the overall uncertainty of the HV measurement was estimated as 0.6\%. 

     \subsection{Single photoelectron station}

\begin{figure}[!hbt]
  \centering  
    \includegraphics[width=0.5\textwidth]{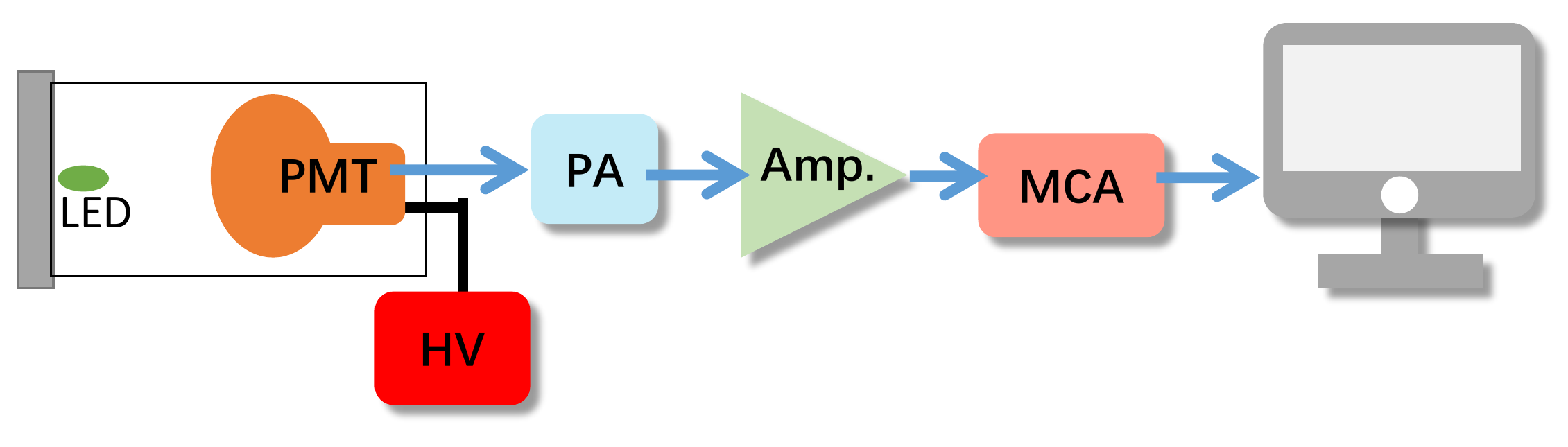}
  \caption{Diagram of SPE station to measure SPE resolution, PV ratio, and DCR. PMT signal was amplified by a preamplifier (PA, CANBERRA Model 2005), a main-amplifier (Amp., CANBERRA Model 2022), and fed into a multichannel analyzer (MCA). The PMT and the LED were in a dark box, while the rest parts were in a room with weak light from the computers' screen and some indicator lamps. The room temperature was controlled by air conditioner at 20$^{\circ}$C}.
  \label{fig:speST}
  \end{figure}

       The SPE station (Fig.~\ref{fig:speST}) was used to measure the SPE spectrum-related parameters (SPE resolution, PV ratio) and DCRs with 0.25~PE and 3.0~PE threshold, respectively. A LED with an appropriate driving voltage provided single 420~nm photons with a distance to PMT of about 15~cm, which fully covered the PMT cathode. The PMT signal was amplified sequentially by two amplifiers and then fed into a 512-channel multichannel analyzer to get the SPE spectrum. Positions of the peak and valley, as well as the FWHM, could be extracted automatically. The SPE resolution \footnote{Defined as FWHM/peak/2.36 in this paper. Some PMT factories and papers use FWHM/peak. For physics study, the $\sigma$ from Gaussian error is more widely used, which is about 2.36 times smaller than FWHM in mathematics.} and the PV ratio could be calculated accordingly. The LED light was turned off when measuring DCR. The DCRs were measured at two thresholds 0.25~PE and 3.0~PE, while the latter was required specifically by JUNO, trying to identify PMTs with large spontaneous light emission. PMTs were kept in the dark box for at least 4 hours before measuring. 
       
\begin{figure}[!hbt]
  \centering  
    \includegraphics[width=0.45\textwidth]{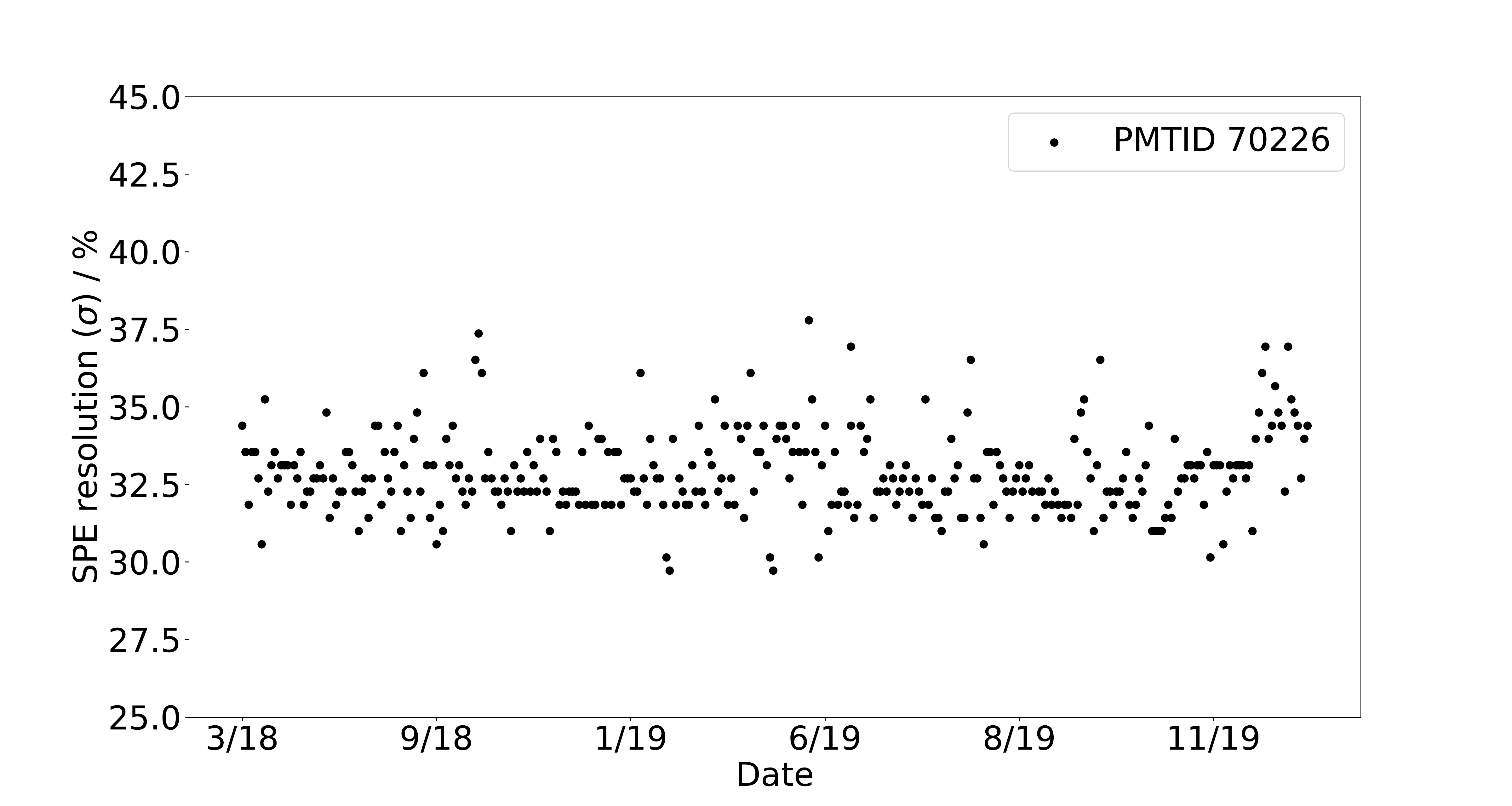}
    \includegraphics[width=0.45\textwidth]{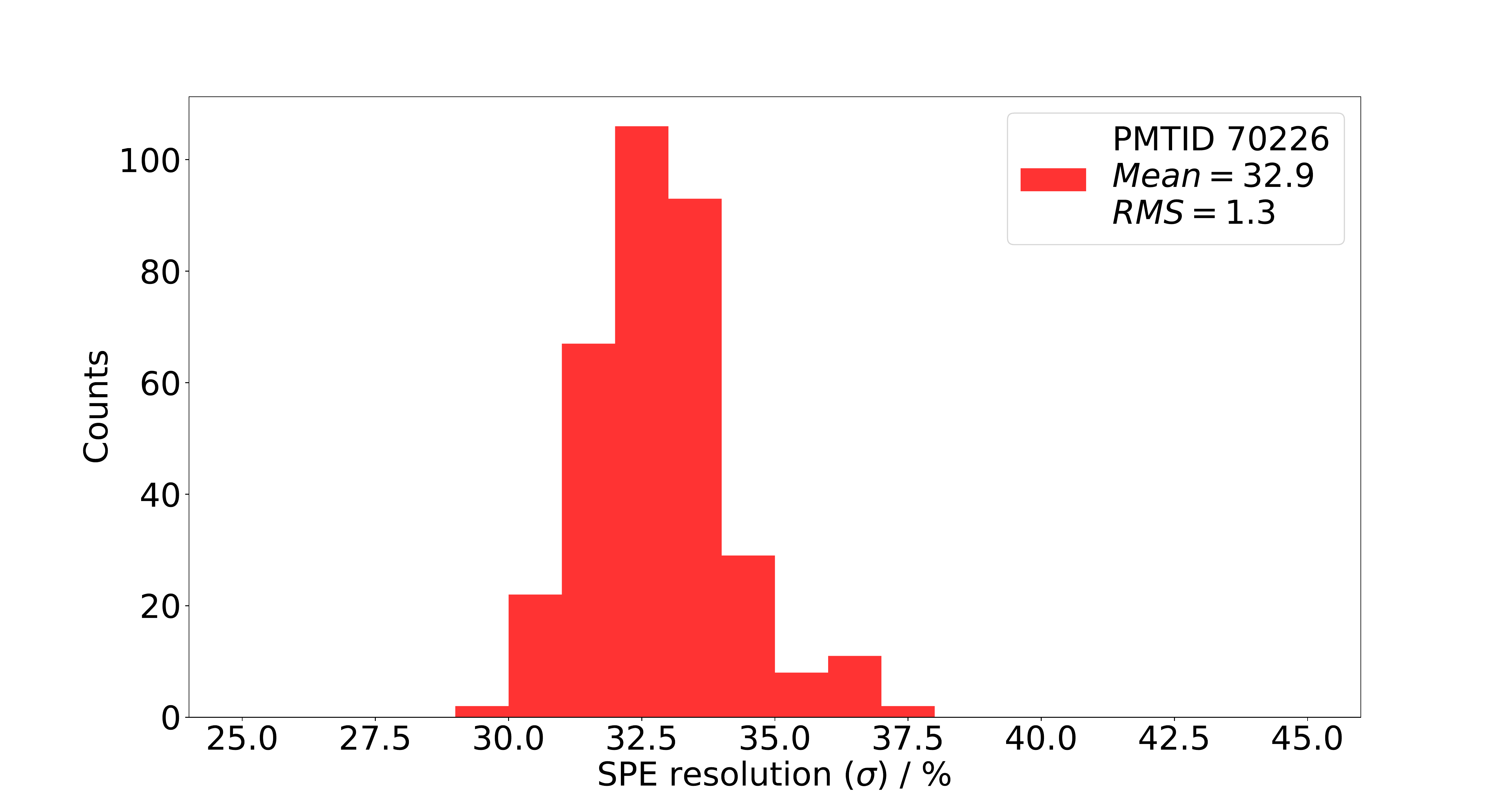}
    \caption{Left: SPE resolution monitoring of the monitor PMT as a function of time. Each point represents one measurement result on working day. Right: Distribution of SPE resolution for the monitor PMT.}
  \label{fig:spe-monitor}
\end{figure}

\begin{figure}[!hbt]
  \centering  
   \includegraphics[width=0.45\textwidth]{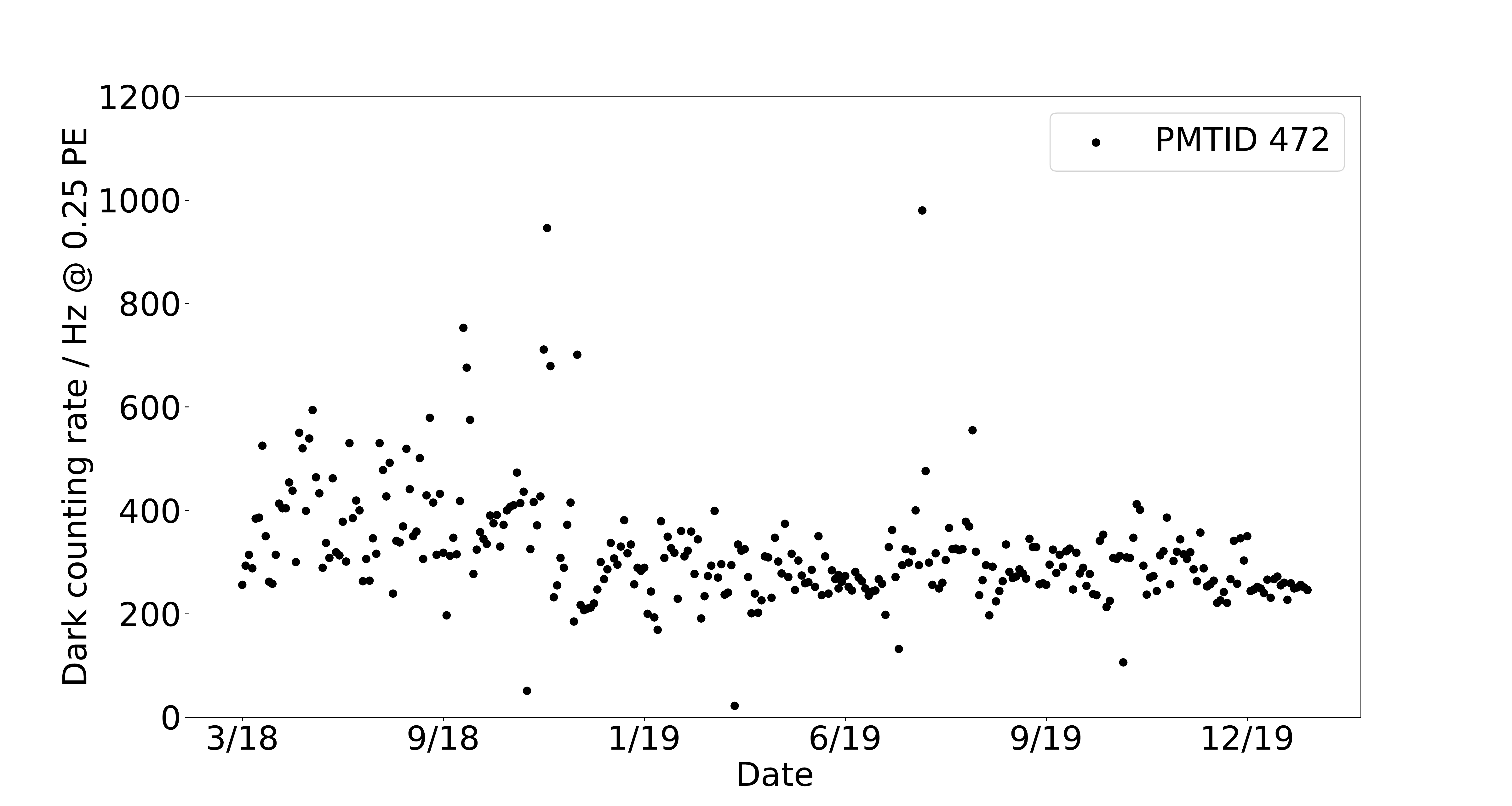}
    \includegraphics[width=0.45\textwidth]{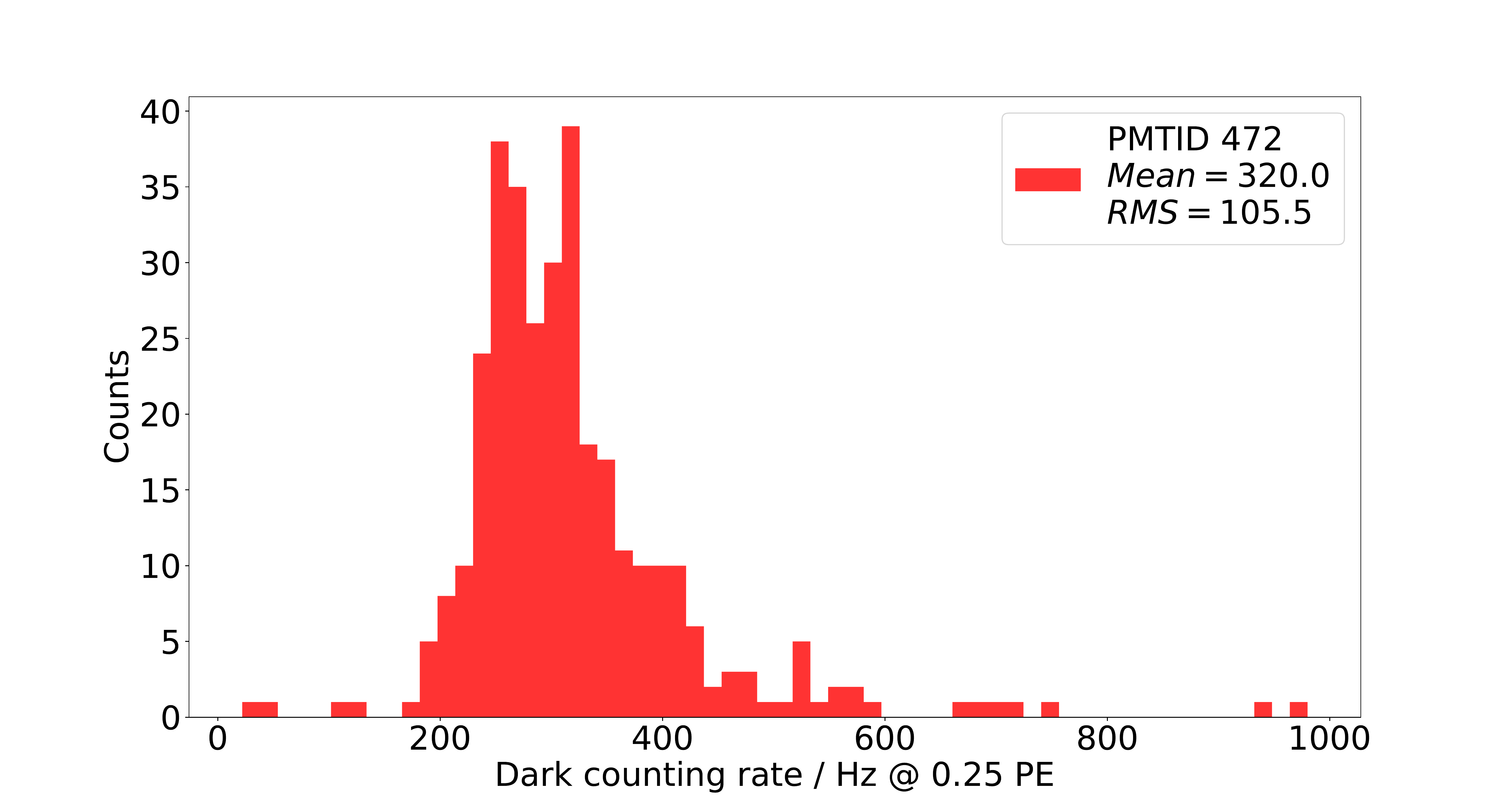}
  \caption{Left: DCR monitoring of the monitor PMT as a function of time. Each point was the measured result in each working day. Right: Distribution of DCR for the monitor PMT.}
  \label{fig:dcr-monitor}
\end{figure}

       There was one PMT selected randomly from the early production to monitor the SPE resolution measurement, as JUNO's requirement. The monitor data of the SPE resolution is shown in Fig.~\ref{fig:spe-monitor}. There was no time-dependent variation but only random fluctuations, showing good stability of the SPE measurement. 
       The relative uncertainty (RMS / Mean) is about 4\%.
        In the factory's standard procedure, another PMT was used just to monitor possible light leakage in the dark box. The DCR monitoring data in Fig.~\ref{fig:dcr-monitor} shows a slow decrease at 0.25~PE threshold in the first several months followed by a period of stability after the PMT was in operation for a longer time. The relative standard deviation 33\% was used to characterize the uncertainty of the DCR measurement.

       \subsection{Transit time spread station}
       
\begin{figure}[!hbt]
           \centering  
           \includegraphics[width=0.5\textwidth]{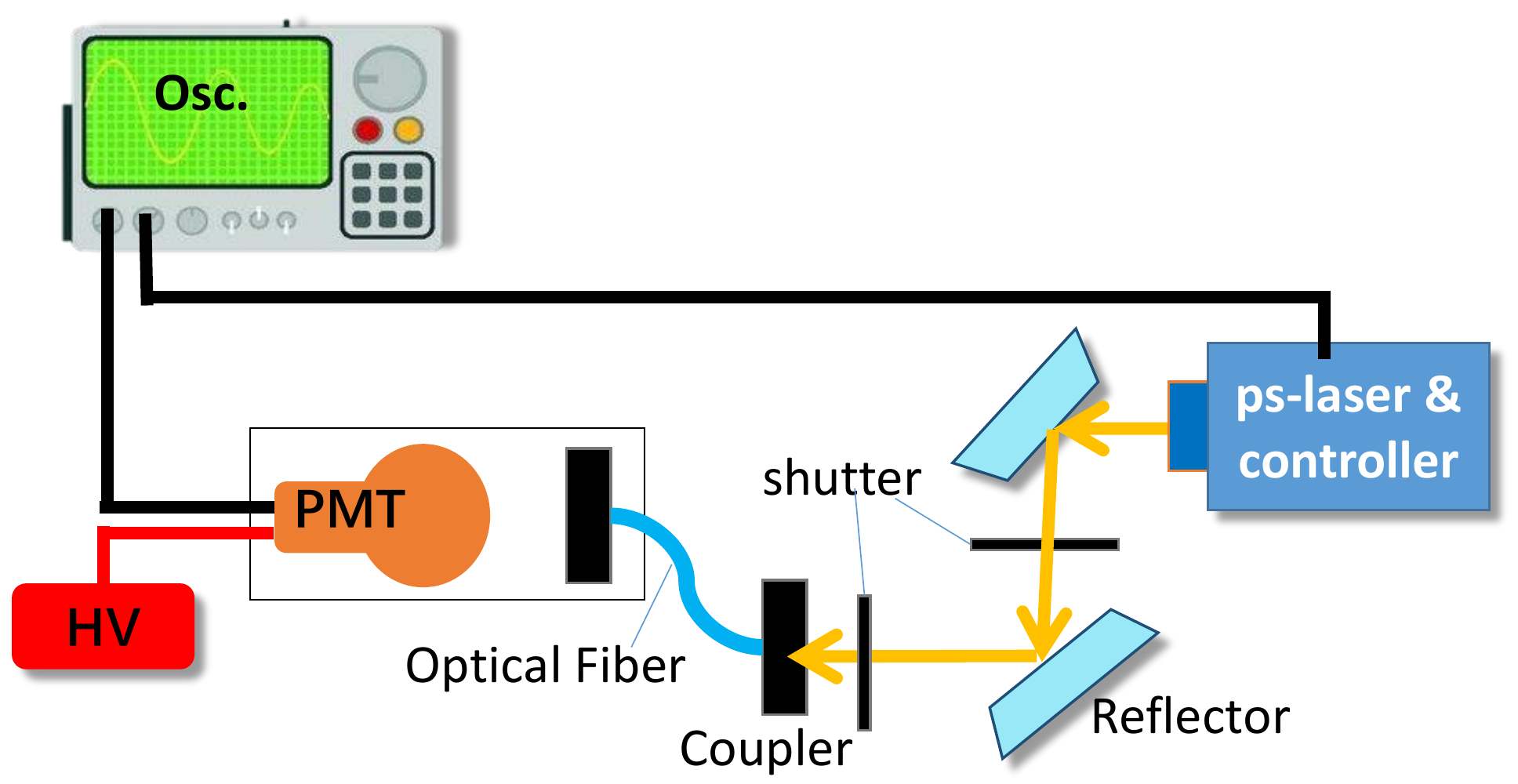}
           \caption{Diagram of the TTS station.}
          \label{fig:TTSST}
        \end{figure}

\begin{figure}[!hbt]
           \centering  
           \includegraphics[width=0.5\textwidth]{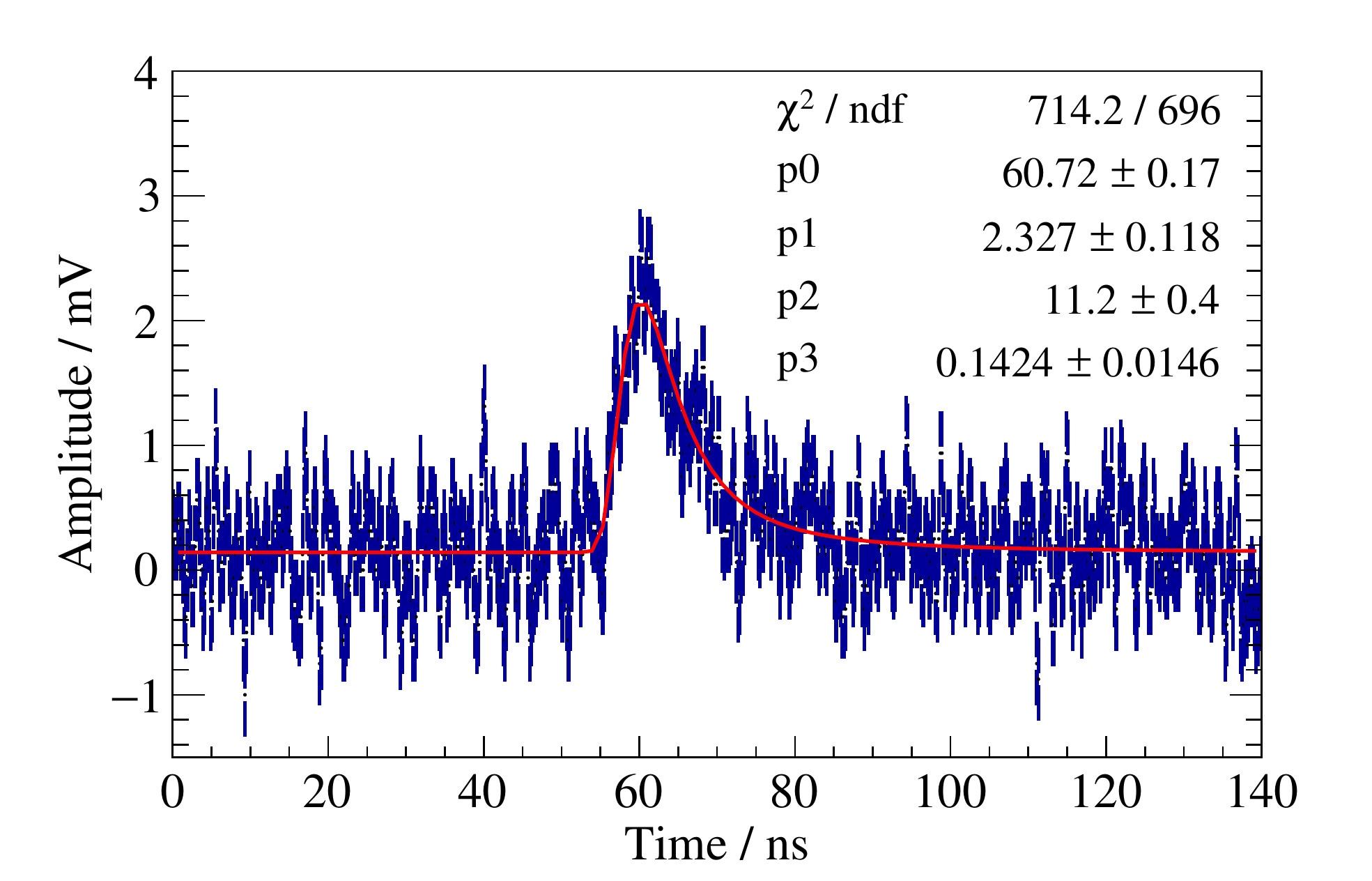}
           \caption{An example of the single PE waveform from one measured PMT, fitting with the Landau function added with a constant baseline. The typical amplitude is between 2 and 3~mV, with the electronics noise smaller than 1~mV. P0 represents the most probable value which is used for timing, p1 and p2 the scale parameters, p3 the baseline.}
          \label{fig:landaufit}
        \end{figure}

      The TTS station shown in Fig.~\ref{fig:TTSST} is a replication of another setup of JUNO~\cite{linan}, which was able to measure not only TTS but also the pre-pulses and the after-pulses. A picosecond laser (405~nm) was used as a light source. The light was reflected and went through a shutter, then into a short plastic optical fiber. There was a divergence angle when the light went out of the fiber into the air and hit the PMT cathode randomly in diameter of $\sim$3~cm. A laser controller was providing two synchronized signals, one to drive the laser pulse and the other to trigger the oscilloscope. The light intensity was adjustable. For the TTS measurement, the average light level was $\sim$0.1~PE, and the time information was extracted by fitting with a Landau distribution (Fig.~\ref{fig:landaufit}). Constant fraction discriminating with low pass filter was also studied, which gave consistent results within 0.1~ns. The timing distribution with sufficient statistics was then fitted with a Gaussian function. In this paper, we use $\sigma$ to express the TTS, which is equal to FWHM/2.36 for a Gaussian distribution. For the pre/after-pulse measurement, the average light level was $\sim$100~PE. Integration of the waveform in the (-90, -10)~ns, (-10, 15)~ns and (0.05, 20)~$\mu$s windows with respect to the peak of the main pulse gave the charge of the pre-pulse $Q_{{\rm pre}}$, main pulse $Q_{{\rm main}}$ and after-pulse $Q_{{\rm after}}$, respectively. The ratio of the pre/after-pulse to the main pulse was calculated as $Q_{{\rm pre}}$/$Q_{{\rm main}}$ and $Q_{{\rm after}}$/$Q_{{\rm main}}$.

\begin{figure}[!hbt]
  \centering  
  \includegraphics[width=0.4\textwidth]{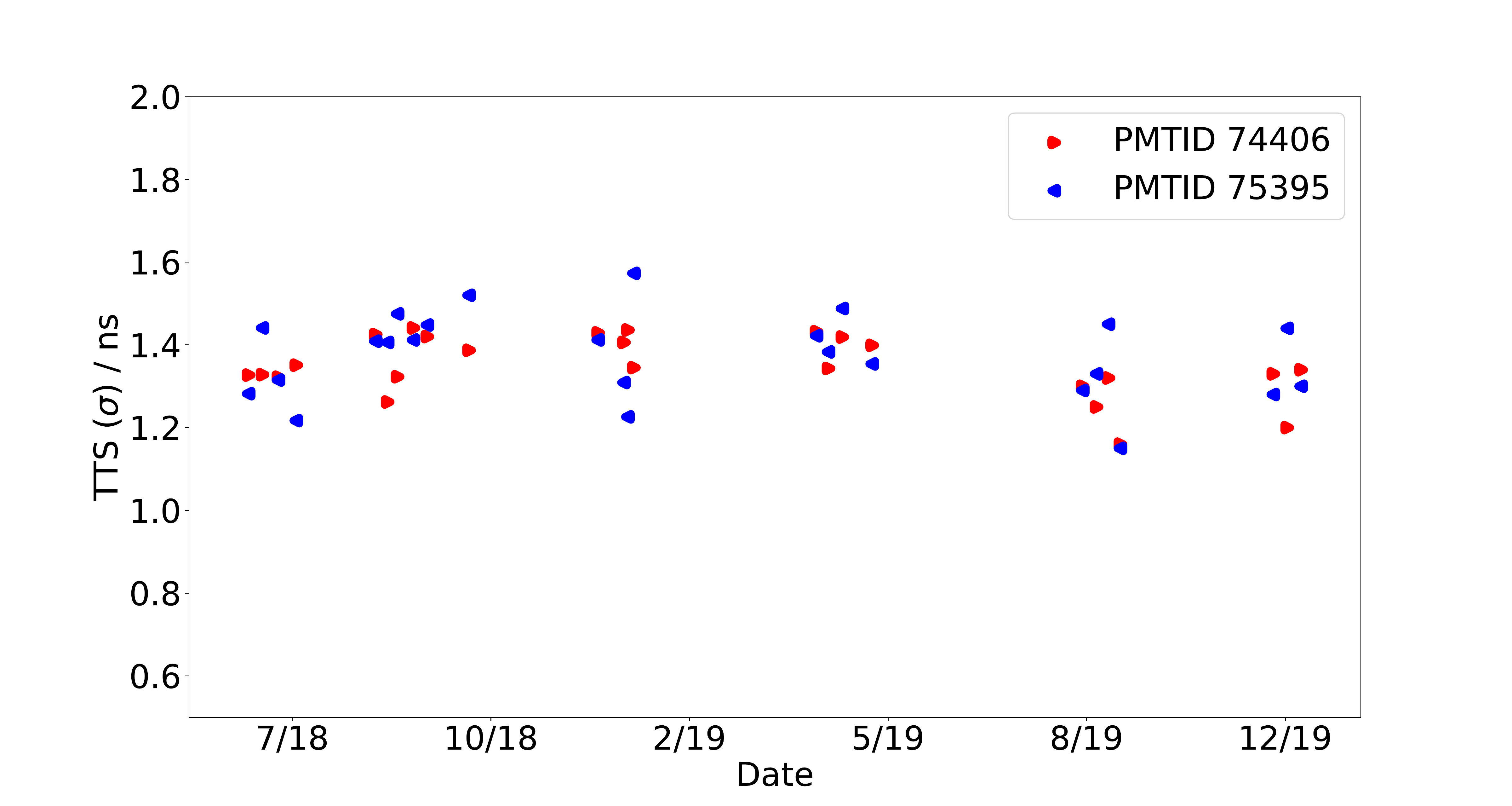}
    \includegraphics[width=0.4\textwidth]{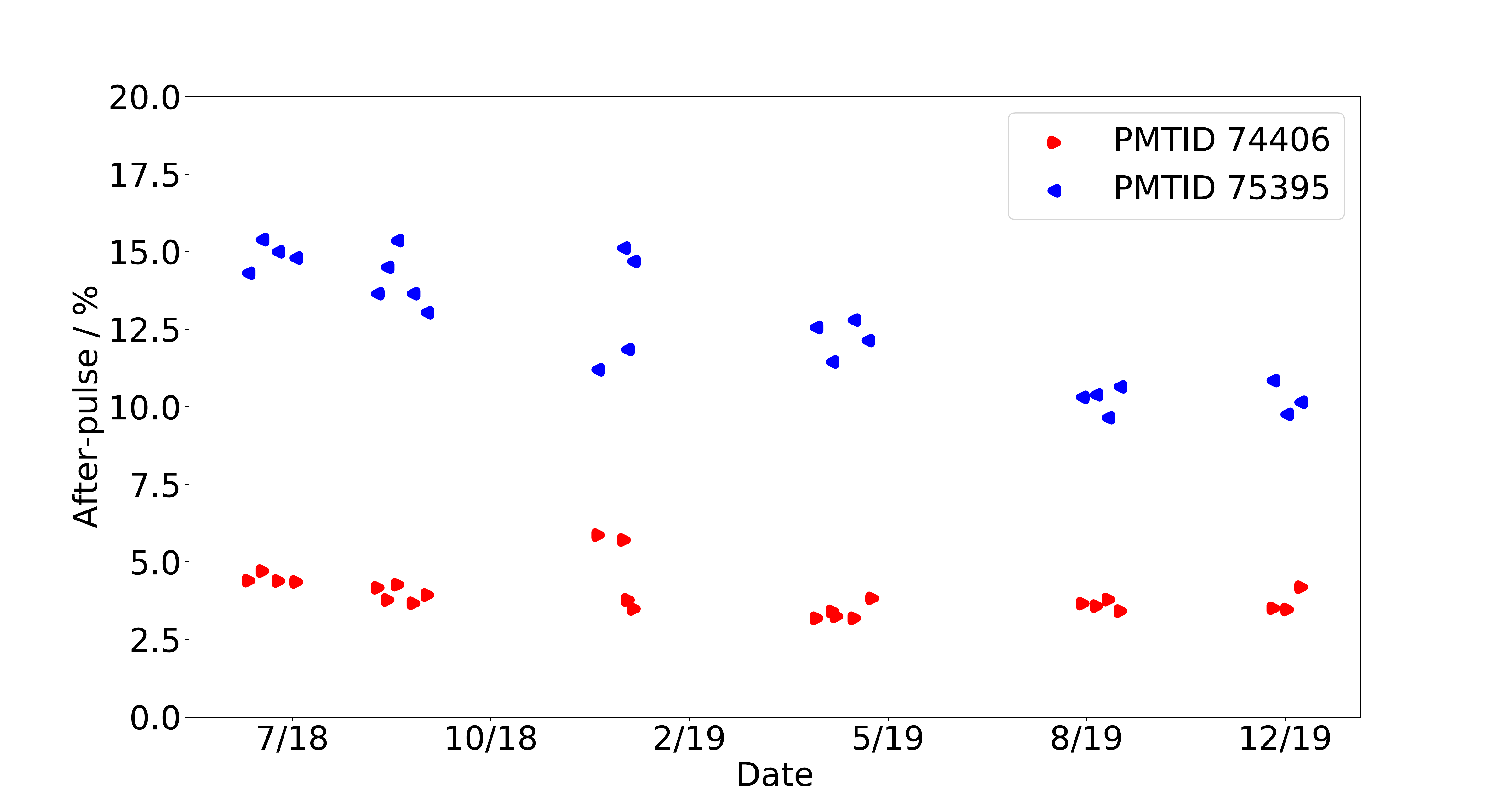}
  \caption{TTS and after-pulse monitoring of two PMTs as a function of time. The data points are grouped reflecting the JUNO onsite testing periods.}
  \label{fig:tts-monitor}
\end{figure}

       There were two PMTs to monitor the long-time stability of the TTS system, as shown in Fig.~\ref{fig:tts-monitor}. Intervals of approximately 3 months can be seen in the plots, corresponding to the onsite testing periods of JUNO. The TTS measurement was very stable with an uncertainty estimated as 10$\%$ based on the standard deviation of all data points. The after-pulse showed a slow decrease in particular for PMT ID~75395, which is a suspect of a continuous ionization of the residual gas molecules in the glass bulb.

  \subsection{Scanning station}

\begin{figure}[!hbt]
  \centering  
        \includegraphics[width=0.5\textwidth]{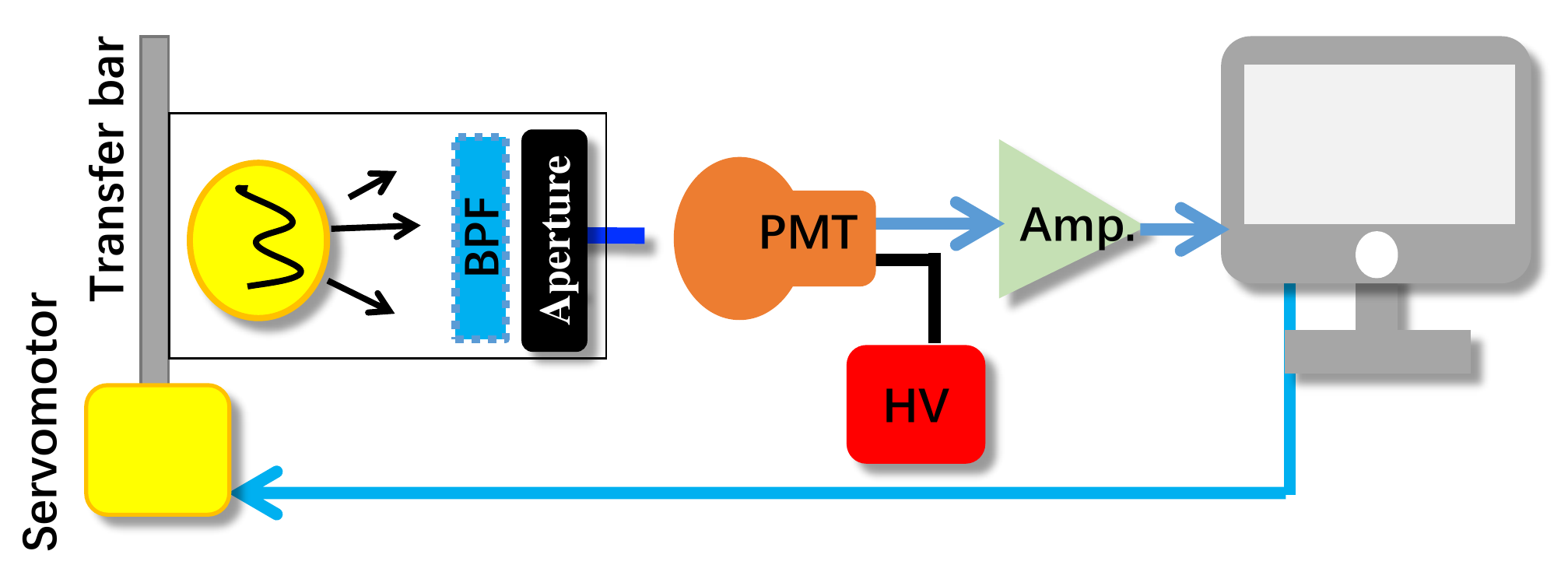}
  \caption{Diagram of the scanning station.}
  \label{fig:scanST}
\end{figure}

\begin{figure}[!hbt]
  \centering  
  \includegraphics[width=0.45\textwidth]{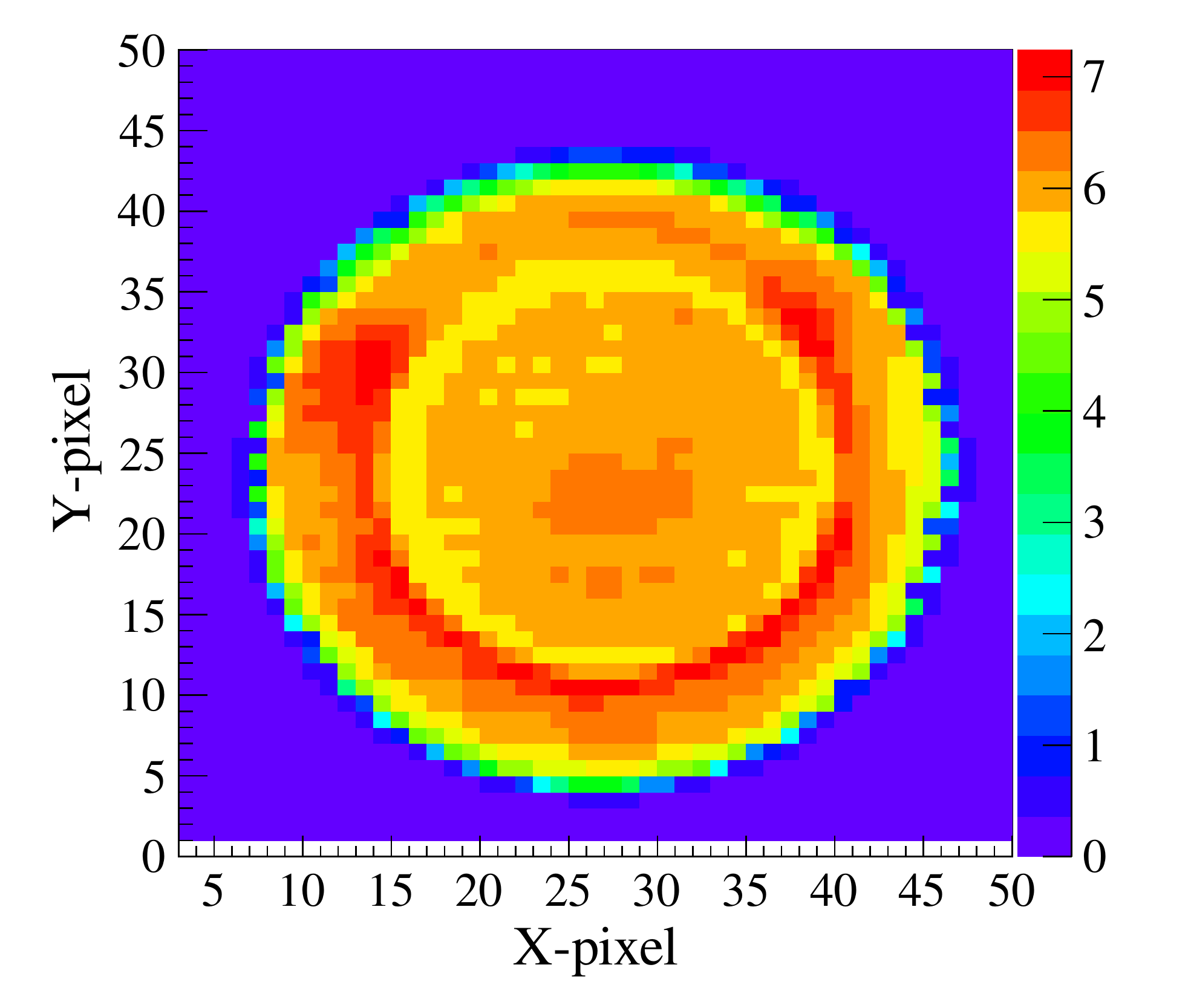}
    \includegraphics[width=0.45\textwidth]{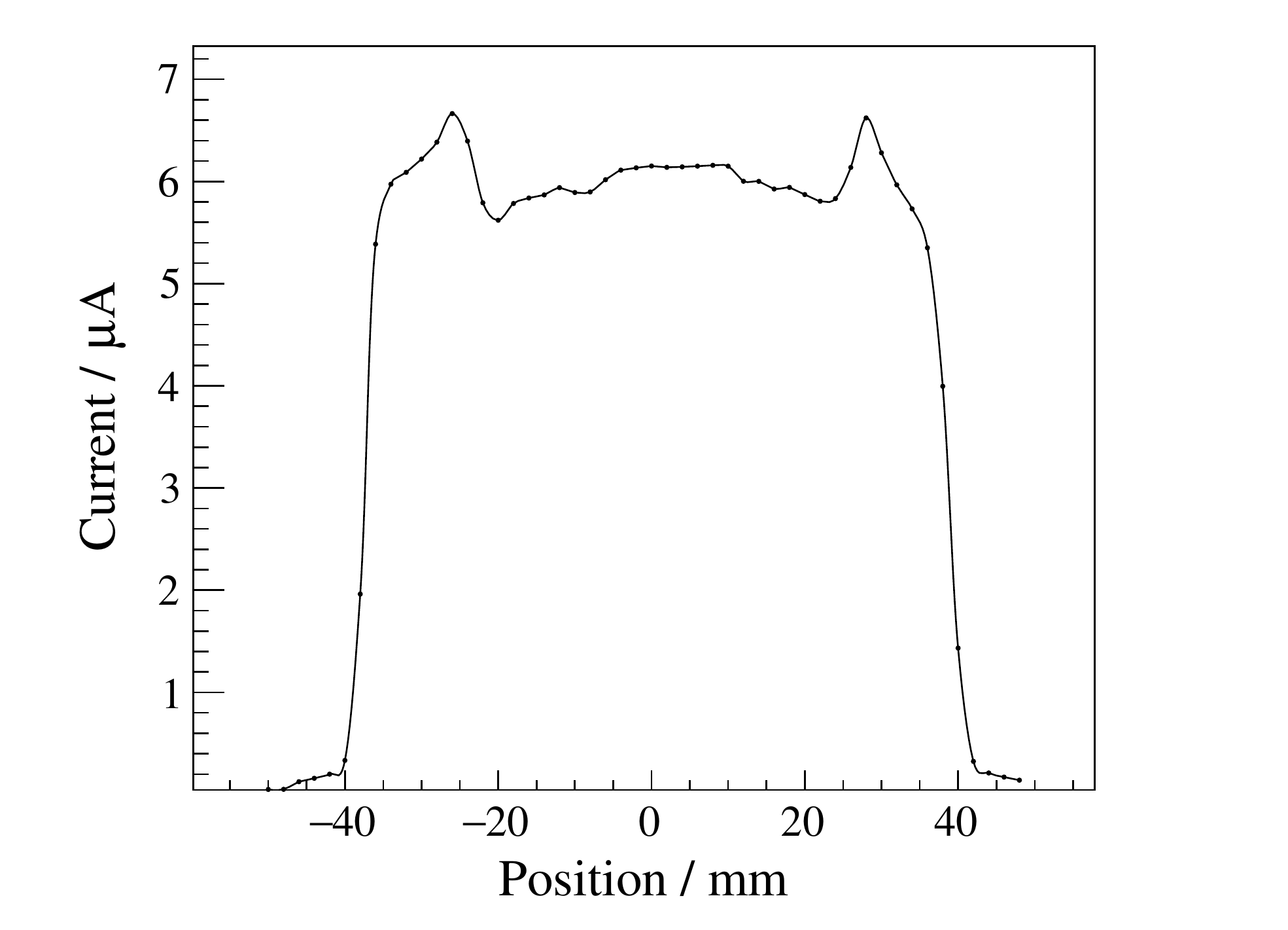}
  \caption{Example of one photocathode scanning. Left: The anode current ($I_{\rm{a}}$) in $\rm{\mu}$A determined for each pixel. Right: The average anode current in the range 20 $<$ Y pixels $<$ 30 as a function of X. The current near the edge of PMT is larger than the central area because of larger incident angle of the light near the edge of the cathode ball and the reflection of the light at the inner side of the lower hemisphere where has an aluminum coating.}
  \label{fig:2Dscan}
\end{figure}

      The scanning station shown in Fig.~\ref{fig:scanST} was required by JUNO to measure the non-uniformity of QE and the effective diameter of the photocathode. A quartz tungsten lamp served as a light source, provided a $\sim$2~mm light spot on the photocathode through a small circular window with a 420~nm filter. The light source, together with the window, could be moved in a 100~mm $\times$100~mm square with 2~mm step size, and thus realized QE scanning in 2,500 pixels covering the photocathode during a testing process. An example of the scanning result of the anode current value is shown in Fig.~\ref{fig:2Dscan}, which portrays the relative changes of QE along the PMT surface by showing the measured anode current ($I_{\rm{a}}$). The two-dimensional projection of the photocathode is clearly demonstrated. There is a ring with higher QE at the edge of the photocathode due to an effect of the glass bulb geometry. The inner area ($\phi$ 60~mm) was used to calculate the QE non-uniformity, expressed as the ratio of the standard deviation to the average. The effective photocathode diameter was determined as an average of diameters determined along the main axes used in the scan. The edges of the photocathode were set at the pixels, for which the anode current drops below 50\% of the average of the inner area.

\subsection{Facilities validation of basic parameters}
\label{sec:validation}

To verify the test facilities at HZC, three parameters that are most important to JUNO were investigated before the mass production: QE, HV, and SPE resolution at a gain of 3$\times$10$^{6}$. Five PMTs were selected randomly and measured by an independent system at the Institute of High Energy Physics~\cite{linan}, and compared with the results by HZC. They were found to be consistent within the uncertainty as shown in Table~\ref{tab:5pmt-comp}.
\begin{table}[!htb]
\caption{Comparison of QE, HV and SPE resolution measurements between JUNO and HZC using the average of 5 PMTs.}\label{tab:5pmt-comp}
 \vspace{0.2cm}
\begin{center}
\begin{tabular}{cccc}\hline
Parameters & QE  & HV  & SPE Res. ($\sigma$) \\
 &   / $\%$ @ 420 nm &  / V @ Gain 3 $\times$ 10$^{6}$ & / $\%$ \\
\hline
JUNO & 24.9 $\pm$ 0.2 & 1074 $\pm$ 5 & 36.6 $\pm$ 1.8\\
HZC  & 24.9 $\pm$ 0.2 & 1070 $\pm$ 5 & 34.1 $\pm$ 1.4\\
\hline
\end{tabular}
\end{center}
\end{table}

\section{Acceptance test and quality assurance}
\label{sec:acc_test}

\subsection{Vendor data statistics and JUNO acceptance strategy}

   All 26,000 3-inch PMTs have been produced, and the 6 parameters from the static station and the SPE station measured by HZC for each PMT. Only PMTs with all of these parameters meeting the requirements were delivered to JUNO. The measured parameters for those (called the vendor data) are shown in Fig. ~\ref{fig:2comp}, where has a cutoff at 900~V and 1,300~V at the HV distribution (900, 1,300) and $<$ 1.8~kHz at the DCR at 0.25~PE distribution. Their averages were also examined and met JUNO's requirements, summarized in Table~\ref{tab:criteria}.

   There were 15 parameters specified by JUNO for the acceptance test as shown in Table~\ref{tab:criteria}. Requirements for these parameters were not only enforced for individual PMTs, but also for the averages of PMT batches. An acceptance test batch contained 3,000 - 6,000 PMTs and there were 7 batches in total. A PMT was rejected if any of the parameters exceeded the limits. 

   The mean value of each batch of PMT production (3,000 PMTs) was also required to pass the mean limit, but it never happened that the mean did not pass the requirements.
   The parameters were divided into four classes according to the test station, test time demands, products quality variation, and the importance to JUNO. Class A parameters were tested by JUNO with 10\% sampling, and 3 parameters (QE, HV, SPE resolution) were compared for further validation if there was a big difference between vendor data and sampling data. Class B and C parameters were only measured by JUNO through random sampling at 3\% and 1\%, respectively, of the PMTs in the whole production. Class D parameters involved destructive measurements and were therefore only done for 3 PMTs. All of the samplings were done by JUNO shifters.

\begin{table}[!htb]
\caption{Summary of the 3-inch PMTs acceptance criteria and test results for different parameters. Results for class A parameters were from 26,000 PMT mean value of vendor data after acceptance measurement introduced in section \ref{sec:acc_procedure}, and other results were from acceptance measurement only. Unless specified, all of the parameters were measured at $3\times10^6$ gain.} \label{tab:criteria}
 \vspace{0.2cm}
\begin{center}
\begin{tabular}{cccccccc}\hline
Parameters &	Class &	\multicolumn{2}{c}{Requirement} & \multicolumn{2}{c}{Test fraction} & Tolerance & Results \\
 & & (limit) & (mean) &  HZC & JUNO & of diff. & (mean) \\
\hline
$\Phi$ (glass bulb) & A & (78, 82)~mm   & -          & 100\% & 10\% & -       & OK      \\
QE@420~nm           & A &	$>$22$\%$   & $>$24$\%$  & 100\% & 10\% & $<$5\%  & 24.9\% \\
High Voltage                  & A & (900,1300)~V  & -          & 100\% & 10\% & $<$3\%  & 1113~V   \\
SPE resolution      & A & $<$45$\%$     & $<$35$\%$  & 100\% & 10\% & $<$15\% & 33.2\% \\
PV ratio            & A & $>$2          & $>$3       & 100\% & 10\% & -       & 3.2    \\
DCR@0.25~PE         & A & $<$1.8~kHz    & $<$1.0~kHz & 100\% & 10\% & -       & 512~Hz \\
DCR@3.0~PE          & A & $<$30~Hz      & -          & 100\% & 10\% & -       & 7.2~Hz \\
TTS ($\sigma$)      & B & $<$2.1~ns       & -          & -     & 3\% & -        & 1.6~ns \\
Pre-pulse           & B & $<$5\%        & $<$4.5\%   & -     & 3\% & -        & 0.5\%  \\
After-pulse         & B & $<$15\%       & $<$10\%    & -     & 3\% & -        & 3.9\%  \\
QE non-uniformity   & B & $<$11\%       & -          & -     & 3\% & -        & 5\%    \\
$\Phi$ (eff. cathode) & B & $>$74~mm      & -          & -     & 3\% & -         & 77.2~mm \\
QE@320~nm           & C &	$>$5\%      & -          & -     & 1\% & -         & 10.2\% \\
QE@550~nm           & C &	$>$5\%      & -          & -     & 1\% & -        & 8.6\%  \\
Aging               & D & $>$200~nA$\cdot$years      & -          & -     & 3 PMTs & - & OK \\
\hline
\end{tabular}
\end{center}
\end{table}

\subsection{Acceptance test for class A parameters in Table \ref{tab:criteria}}
\label{sec:acc_procedure}

\begin{figure}[!hbt]
  \centering  
  \includegraphics[width=0.45\textwidth]{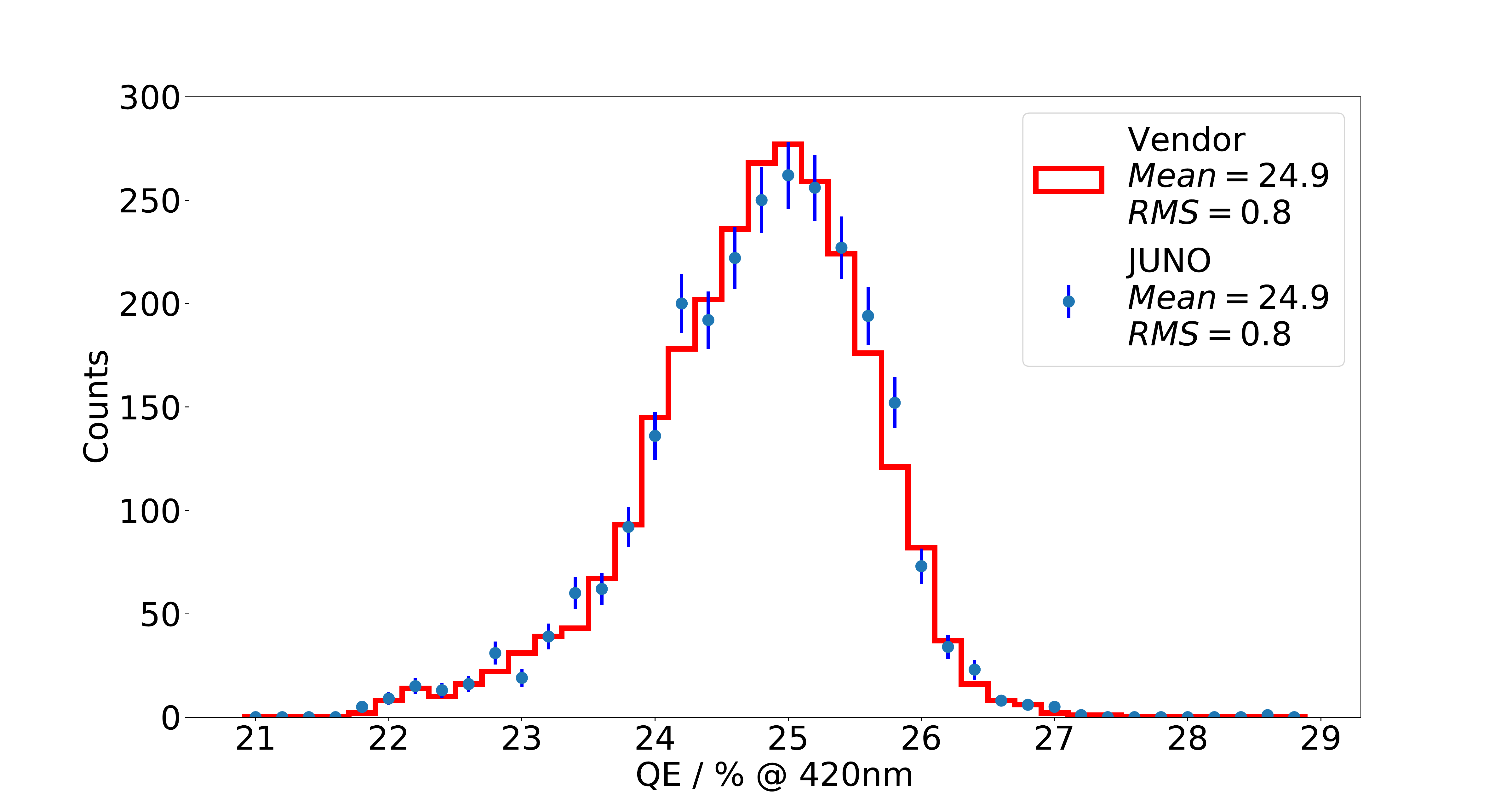}
  \includegraphics[width=0.45\textwidth]{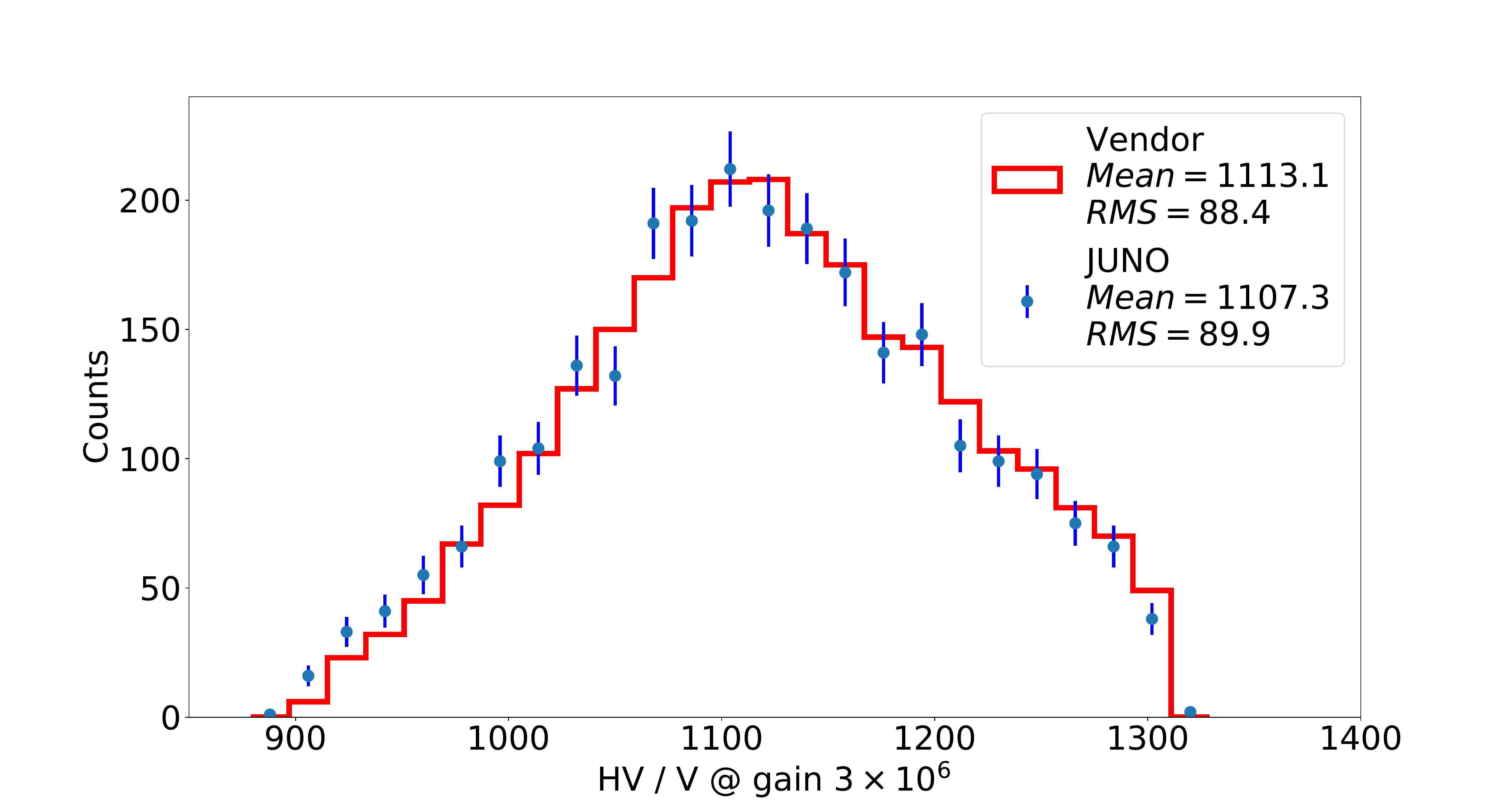}
  \includegraphics[width=0.45\textwidth]{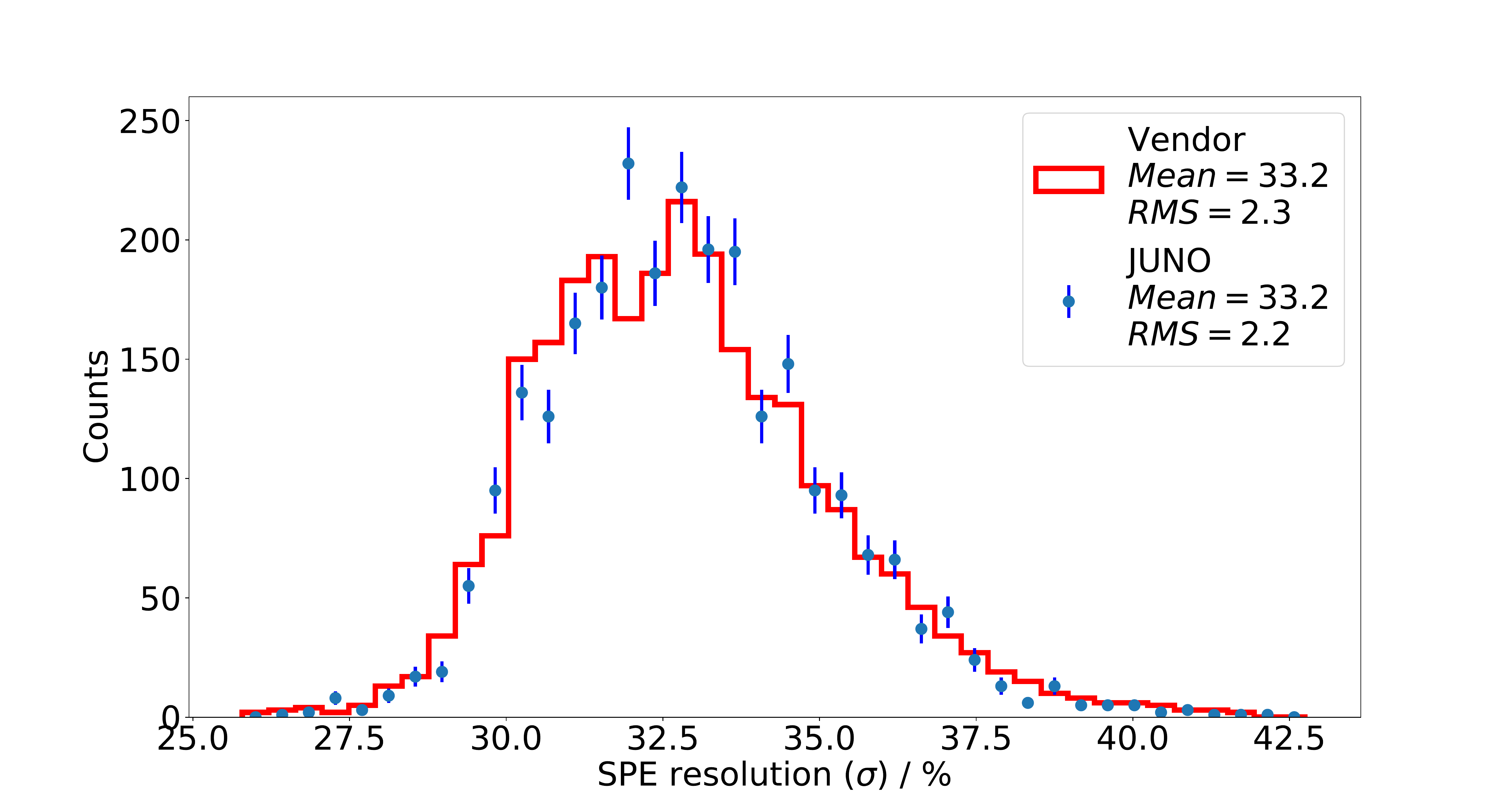}
  \includegraphics[width=0.45\textwidth]{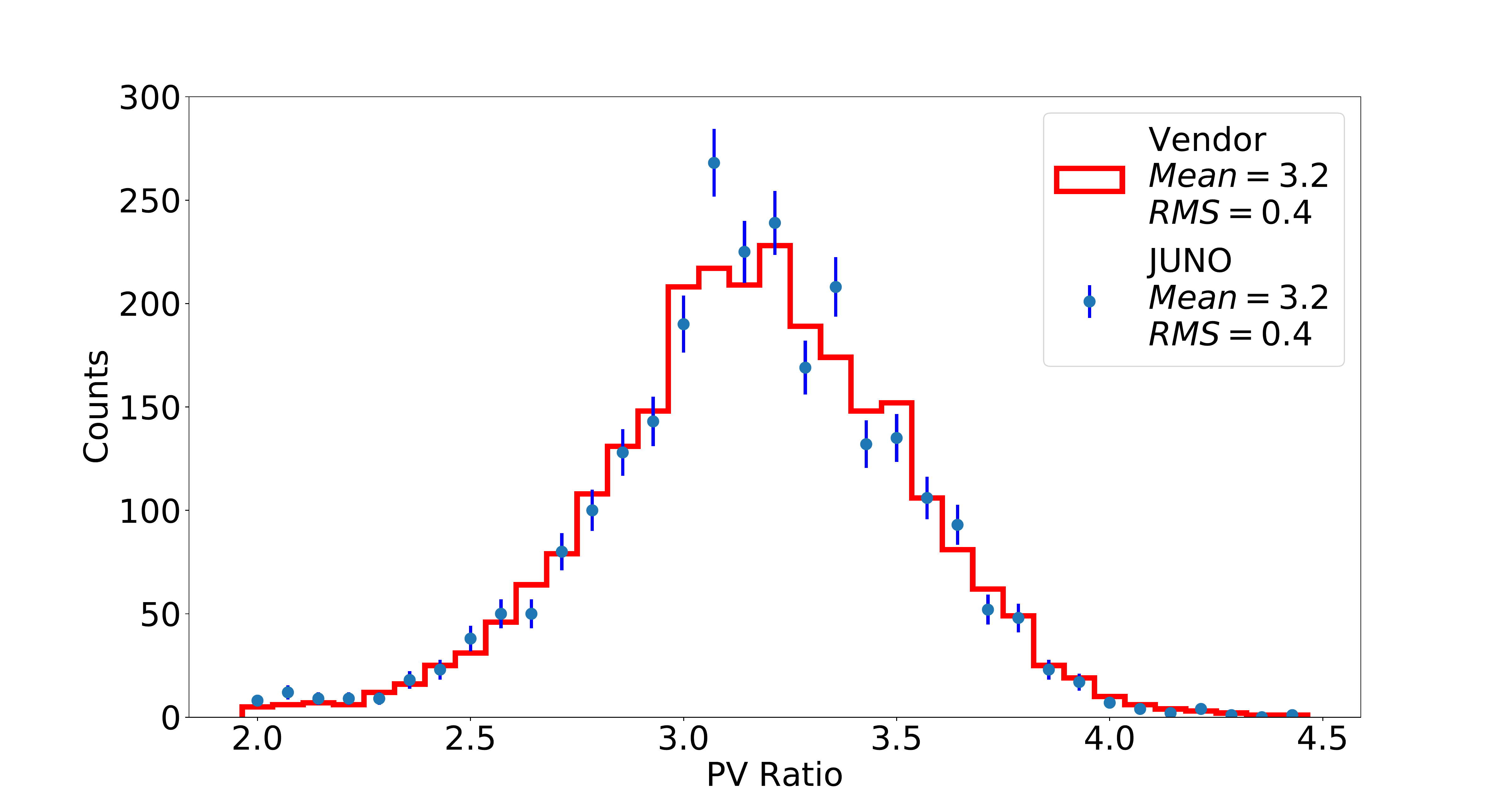}
  \includegraphics[width=0.45\textwidth]{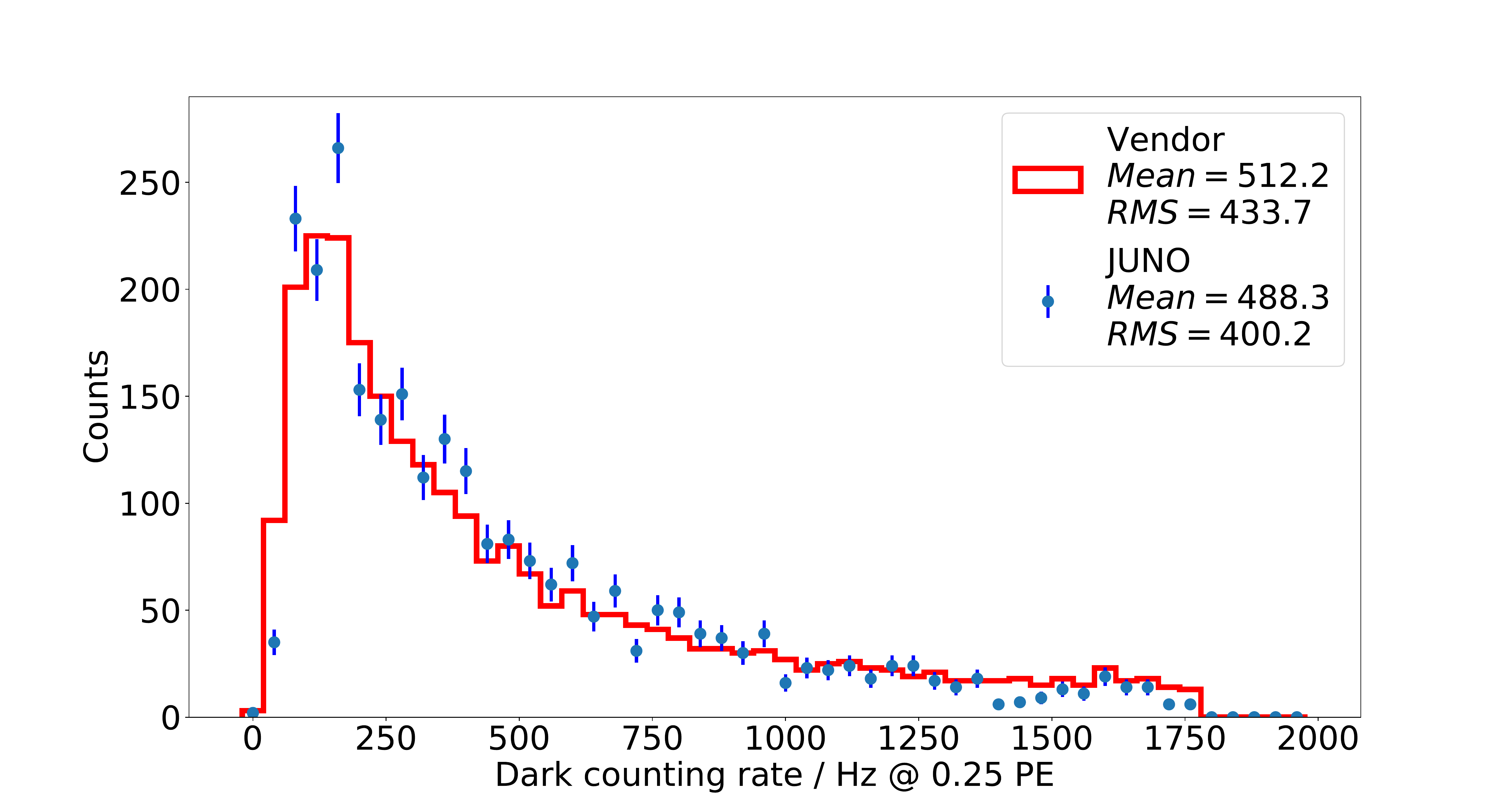}
  \includegraphics[width=0.45\textwidth]{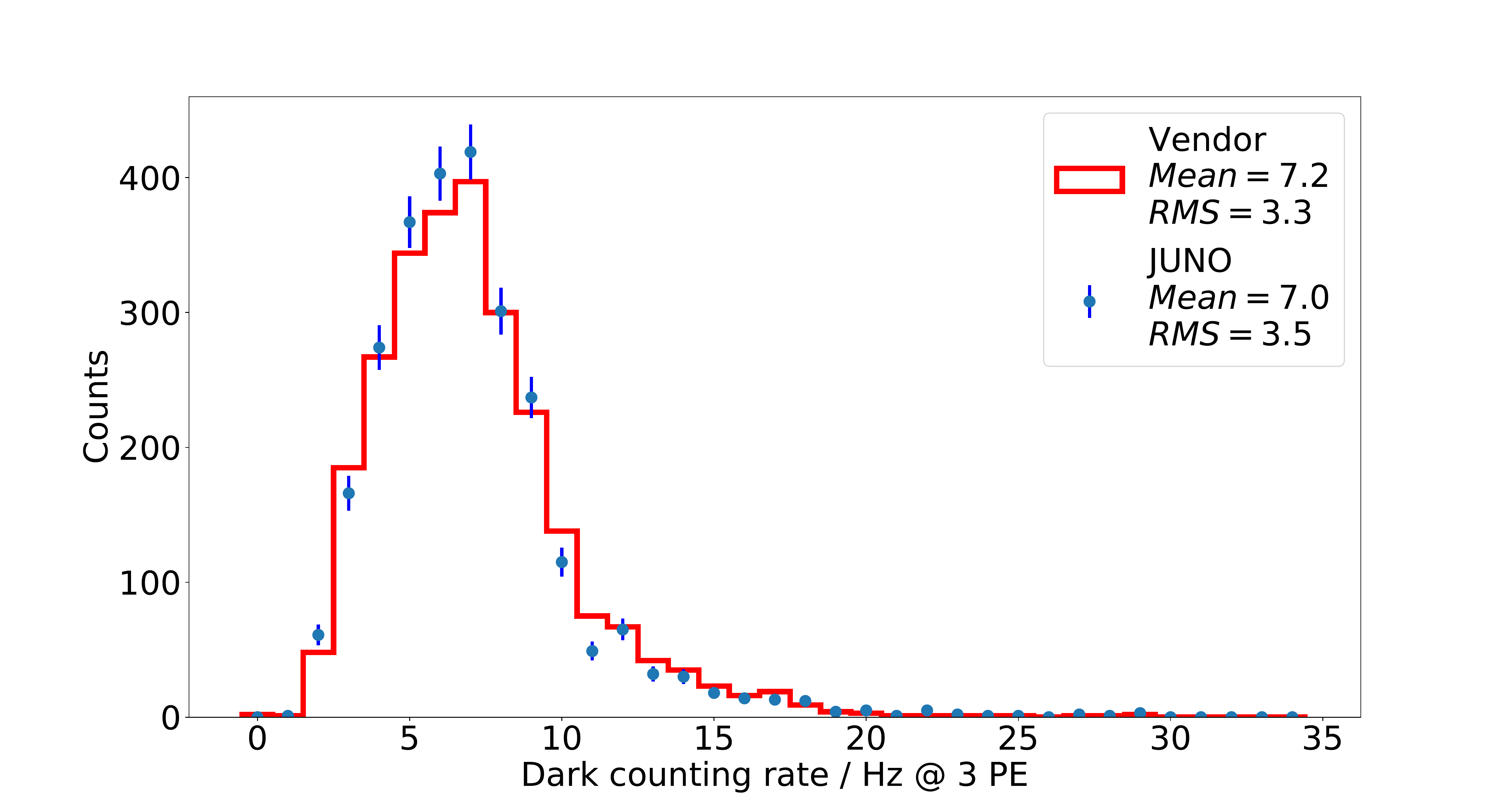}
  \caption{The PMT sampling test results (2,600 PMTs) for class A parameters and comparison with vendor data (26,000 PMTs) after normalization.}
  \label{fig:2comp}
\end{figure}

   There were 7 parameters contained in class A: the diameter of the glass bulb, QE, HV, SPE resolution, PV ratio, DCRs at 0.25~PE, and 3.0~PE threshold. 150 PMTs were defined as a sub-batch since 75 PMTs were packaged in one box. 10\% of them were randomly selected by the JUNO shifter. The diameter was examined first by two rings with inner diameters of 78~mm and 82~mm. After that, the sampled PMTs were delivered to the HZC worker to test at the static station and the SPE station, and the results were sent back to the JUNO shifter. If any parameter was found to exceed the limitation, this PMT was measured again. If the second test gave the same result, this PMT was rejected and replaced with a new one. Among all 2,600 PMTs selected for class A parameter acceptance measurements, only 3 were rejected at this step, one with HV lower than 900~V, one with DCR at 0.25~PE larger than 1.8~kHz, and one with DCR at 3.0~PE larger than 30~Hz. The sampling test results are compared with the vendor data in Fig.~\ref{fig:2comp}, obtaining good consistency.

\begin{figure}[!hbt]
  \centering  
  \includegraphics[width=0.3\textwidth]{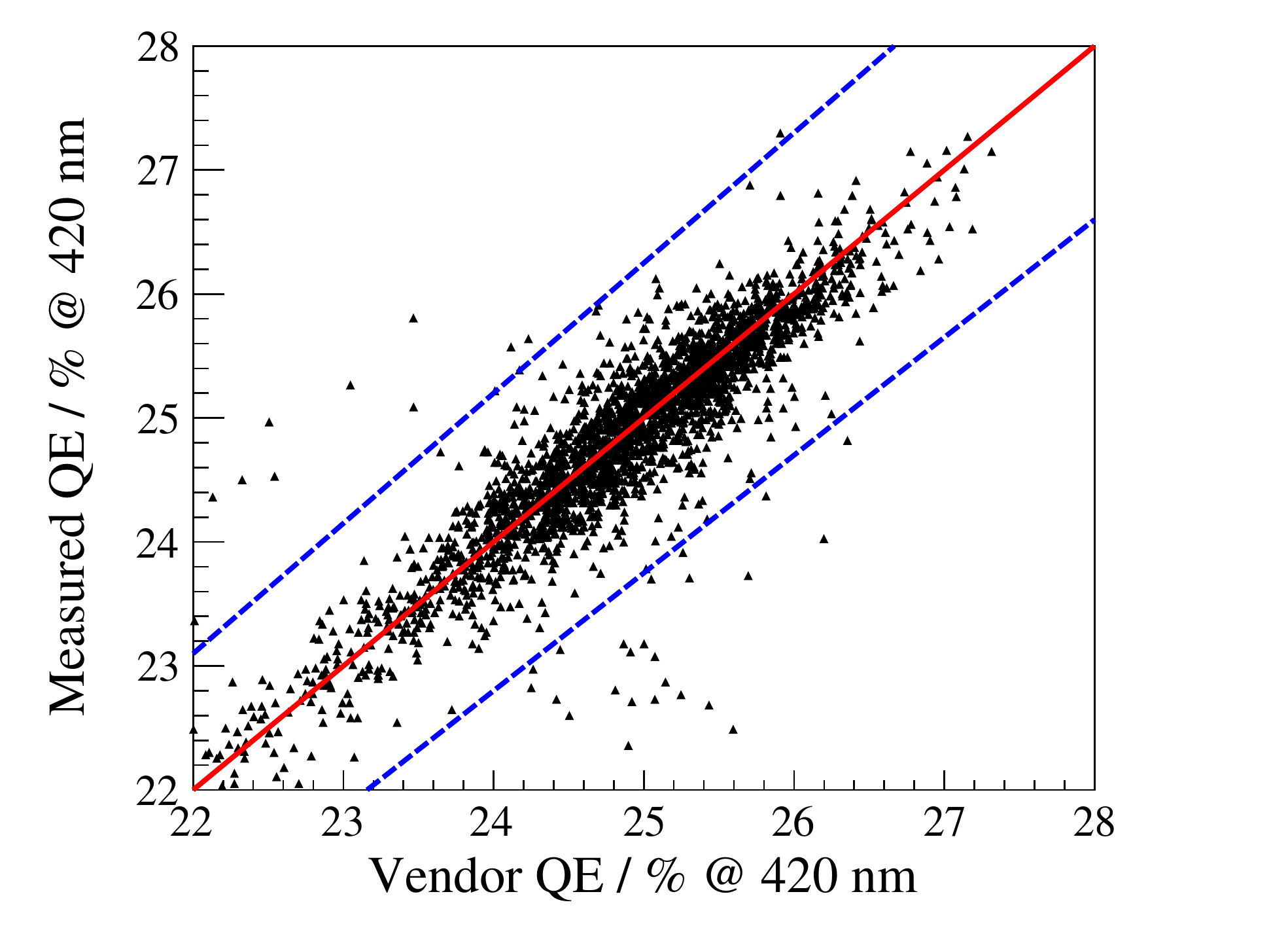}
    \includegraphics[width=0.3\textwidth]{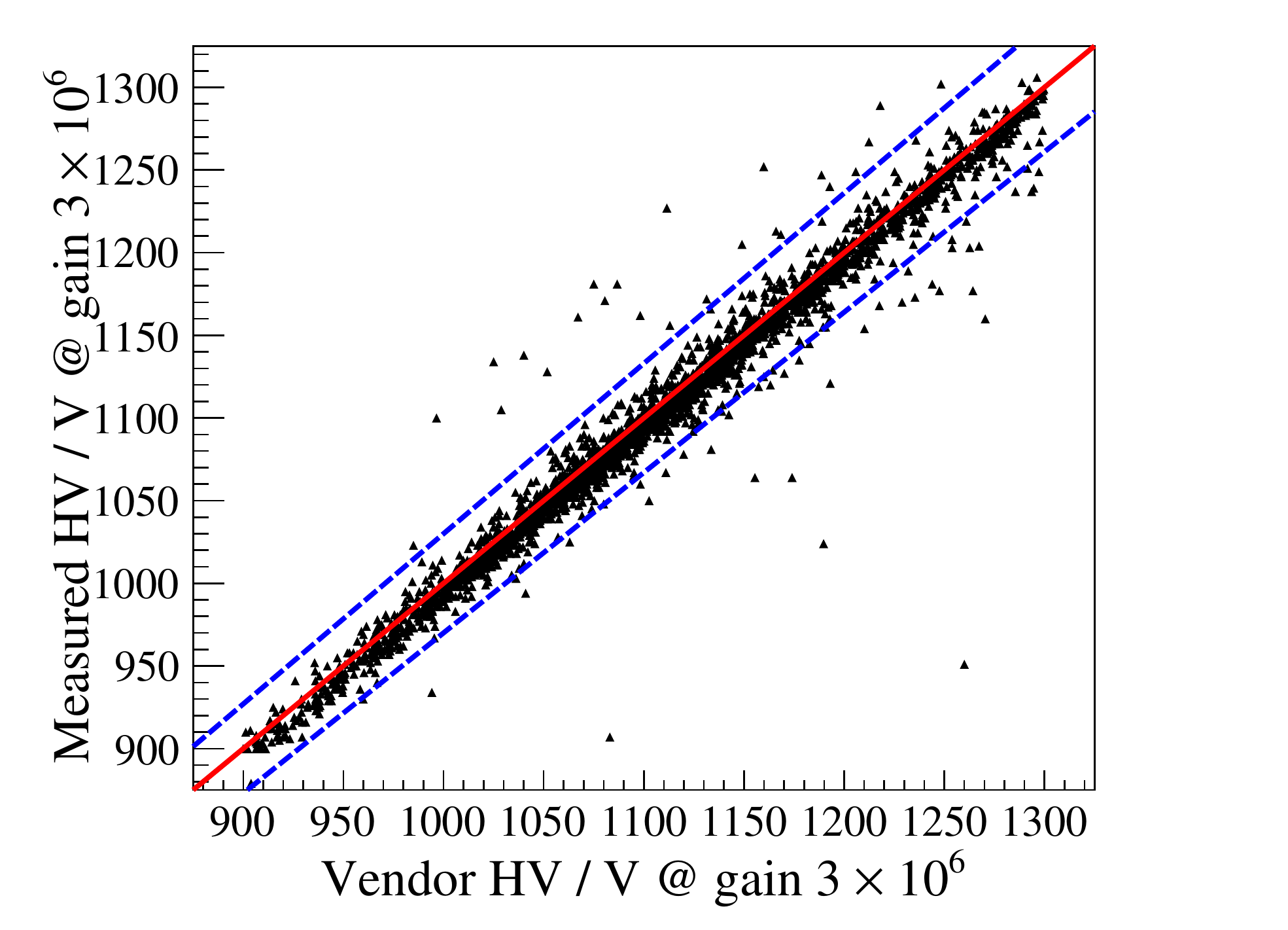}
      \includegraphics[width=0.3\textwidth]{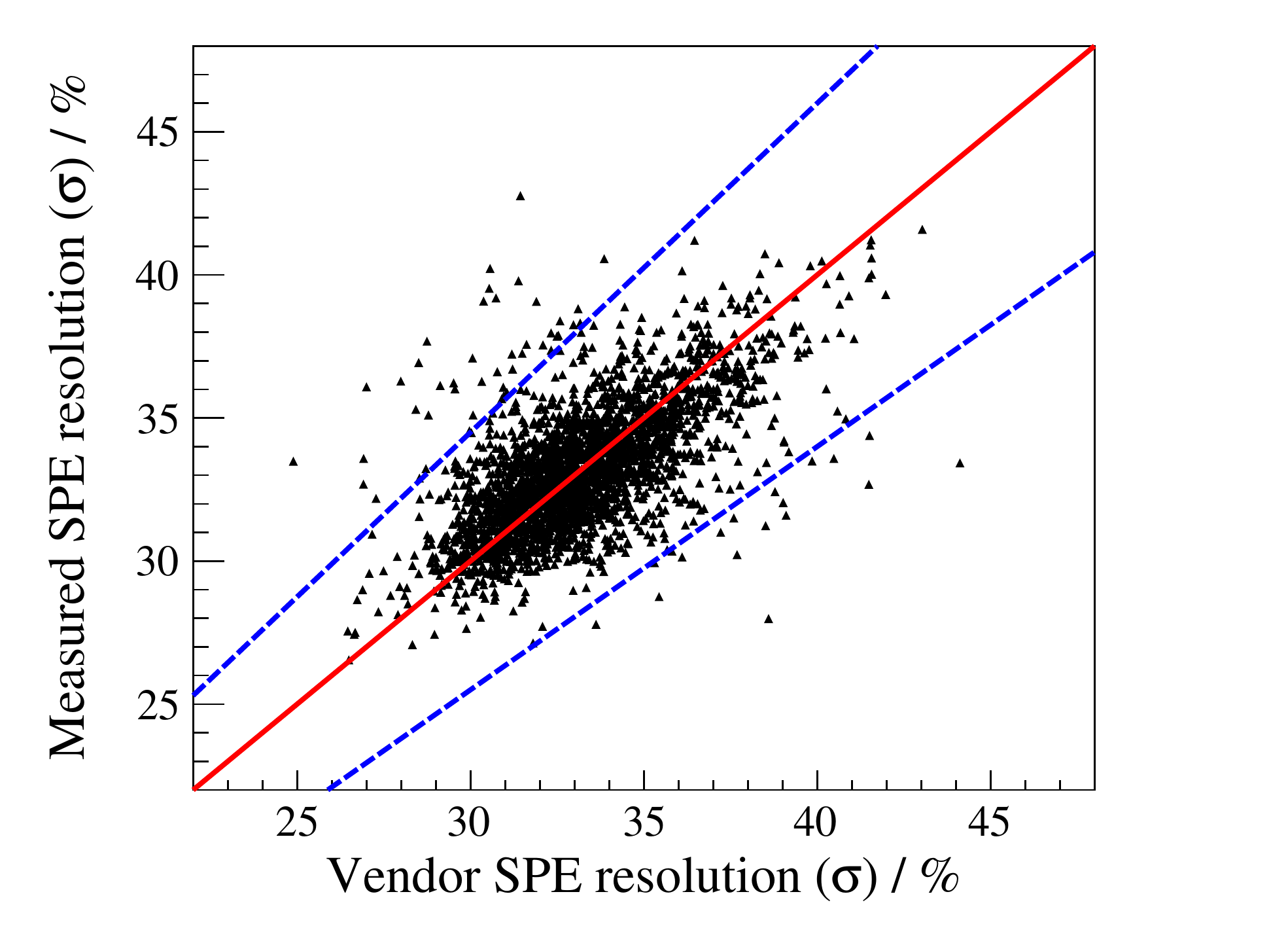}
  \caption{A comparison of the sampling test result with vendor data for the same 2,600 PMTs, for QE, HV and SPE resolution, respectively. The red lines are the proportional function. The blue dash lines are the maximum and minimum tolerance ranges.}
  \label{fig:3par-comp}
\end{figure}

   For QE, HV, and SPE resolution, the difference between the sampling test results and the vendor data was required to be smaller than a tolerance, defined as 5\%, 3\%, and 15\% for each single PMT, respectively, based on the 2-year stability results of the test stations reported in Sec.~\ref{sec:test_station}. These tolerances corresponded to 4-6~$\sigma$ to allow the normal fluctuation to be accepted. Only exceptions, such as a sudden change of the test system performance, an unstable PMT, or a human mistake was expected to be caught. Once a big difference was found, a second test was done for the problematic PMT. If the second result was consistent with the vendor data, this PMT would be accepted. If the two rounds of sampling test agreed with each other but were far from the vendor data, this PMT would be also accepted but the vendor data would be changed to the new one. In the worst case that all of these 3 tests were very different, this PMT would be rejected. Only one PMT was rejected at this step because of unstable QE. The comparison of the first sampling test result with the vendor data for the same 2,600 PMTs is shown in Fig.~\ref{fig:3par-comp}, with the tolerances range depicted. The fractions of PMTs out of tolerances were 1.6\%, 2.7\%, and 2.4\% for QE, HV, and SPE, respectively, and the majority agreed with the vendor data after re-testing.

\begin{figure}[!hbt]
  \centering  
  \includegraphics[width=0.9\textwidth]{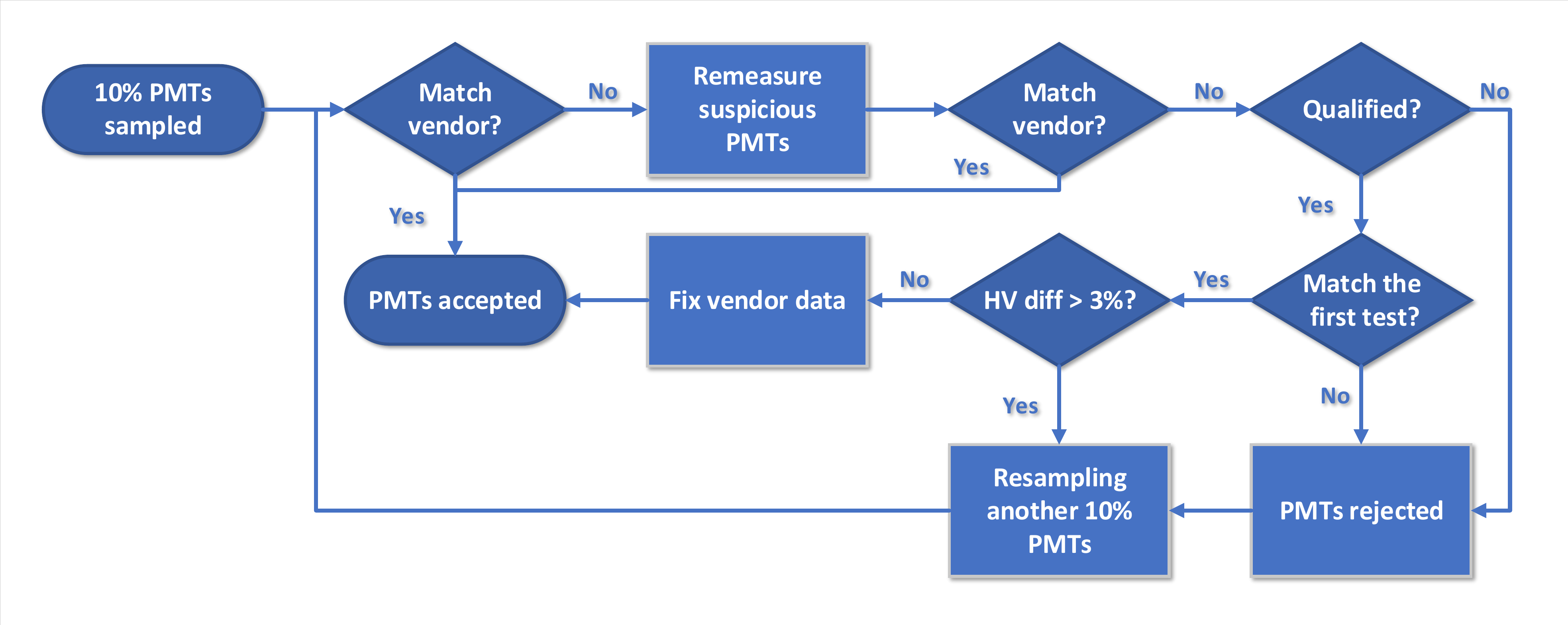}
  \caption{The flow chart of sampling test for HV parameter. The procedure was operated in each sub-batch PMTs (150 PMTs).}
  \label{fig:procedure}
\end{figure}

   In JUNO, groups of sixteen 3-inch PMTs will be powered with one single HV channel. Therefore, the working HV measurement was required to be more reliable than other parameters to ensure that the gains of all PMTs in each group are as close as possible. Once a PMT was rejected because of HV, we re-sampled another 15 PMTs (10\%) in the same sub-batch and repeated the above procedure. The full test procedure is shown in Fig.~\ref{fig:procedure}, using HV as the most complicated example.

\begin{figure}[!hbt]
  \centering  
  \includegraphics[width=0.8\textwidth]{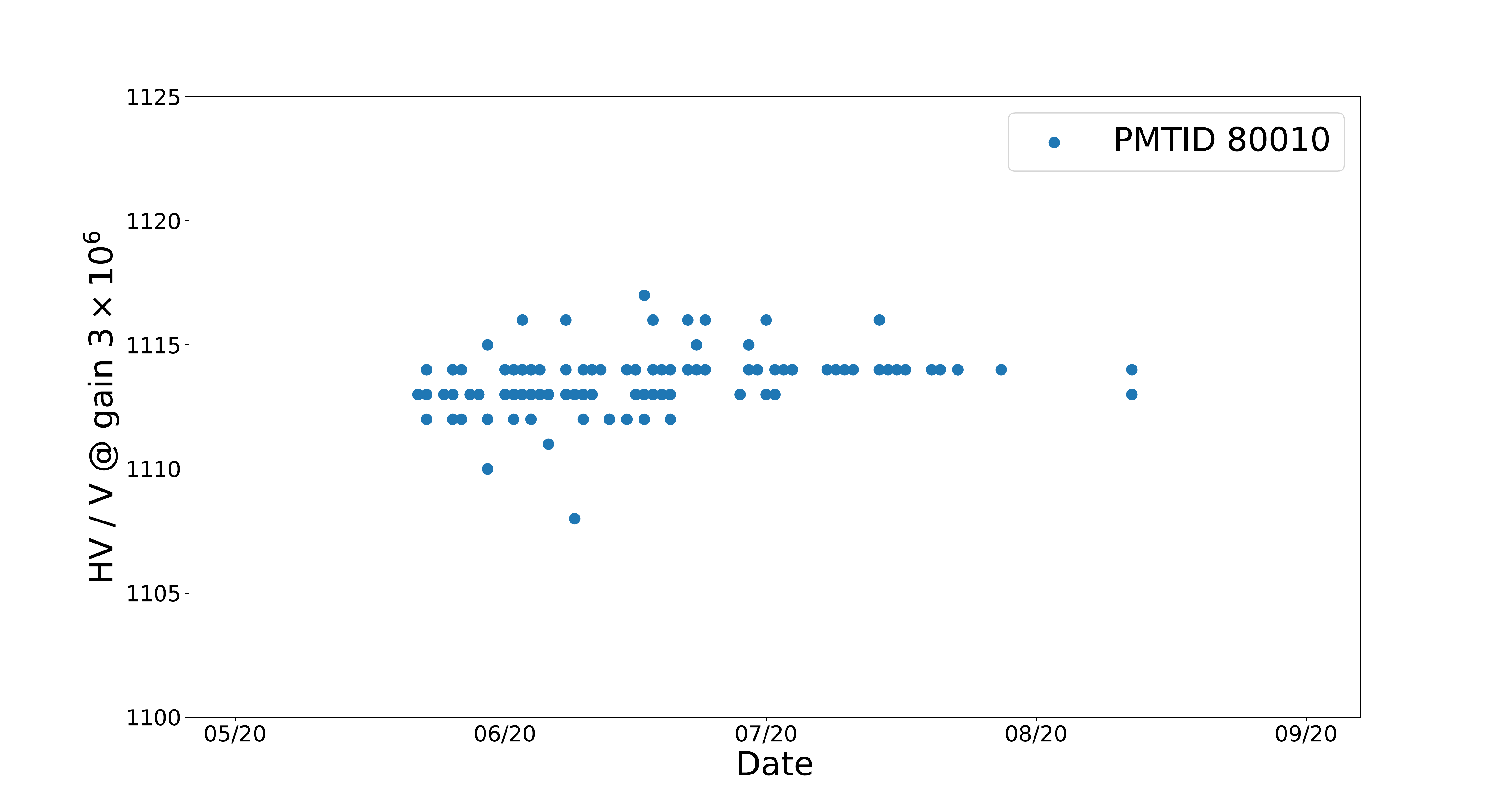}
  \caption{HV monitoring during re-testing of the first 10,000 PMTs. Each point shows the measured result of a randomly selected PMT (ID 80010) in each working day.}
  \label{fig:re-test-HV-monitor}
   \end{figure}

   Because of the large fluctuation of the HV measurement before August 2018, as indicated in Fig.~\ref{fig:HV-monitor}, the first 10,000 PMTs produced and tested in that period were tested again in 2020 with a new system, whose principle was the same as the static station (Fig.~\ref{fig:staticST}), but with better control of the noise. One JUNO PMT was randomly selected and tested in this system every working day to monitor the stability, as shown in Fig.~\ref{fig:re-test-HV-monitor}. The variations were found to be within $\pm$~5~V.

\subsection{Acceptance test for class B and C parameters in Table \ref{tab:criteria}}

   The ratio of PMTs tested for parameter classes B and C were 3\% and 1\%, respectively, both sampled by the JUNO shifter from those 15 PMTs (10\%) which were tested in class A acceptance test. Similarly, the class C sample was fully contained in the class B sample, resulting in 1\% of all PMTs being fully characterized. All of these parameters were required to be within the limits or a second test was done. PMTs with two failed tests were rejected and a re-sampling of 3\% or 1\% in this sub-batch was required.

\begin{figure}[!hbt]
  \centering  
  \includegraphics[width=0.3\textwidth]{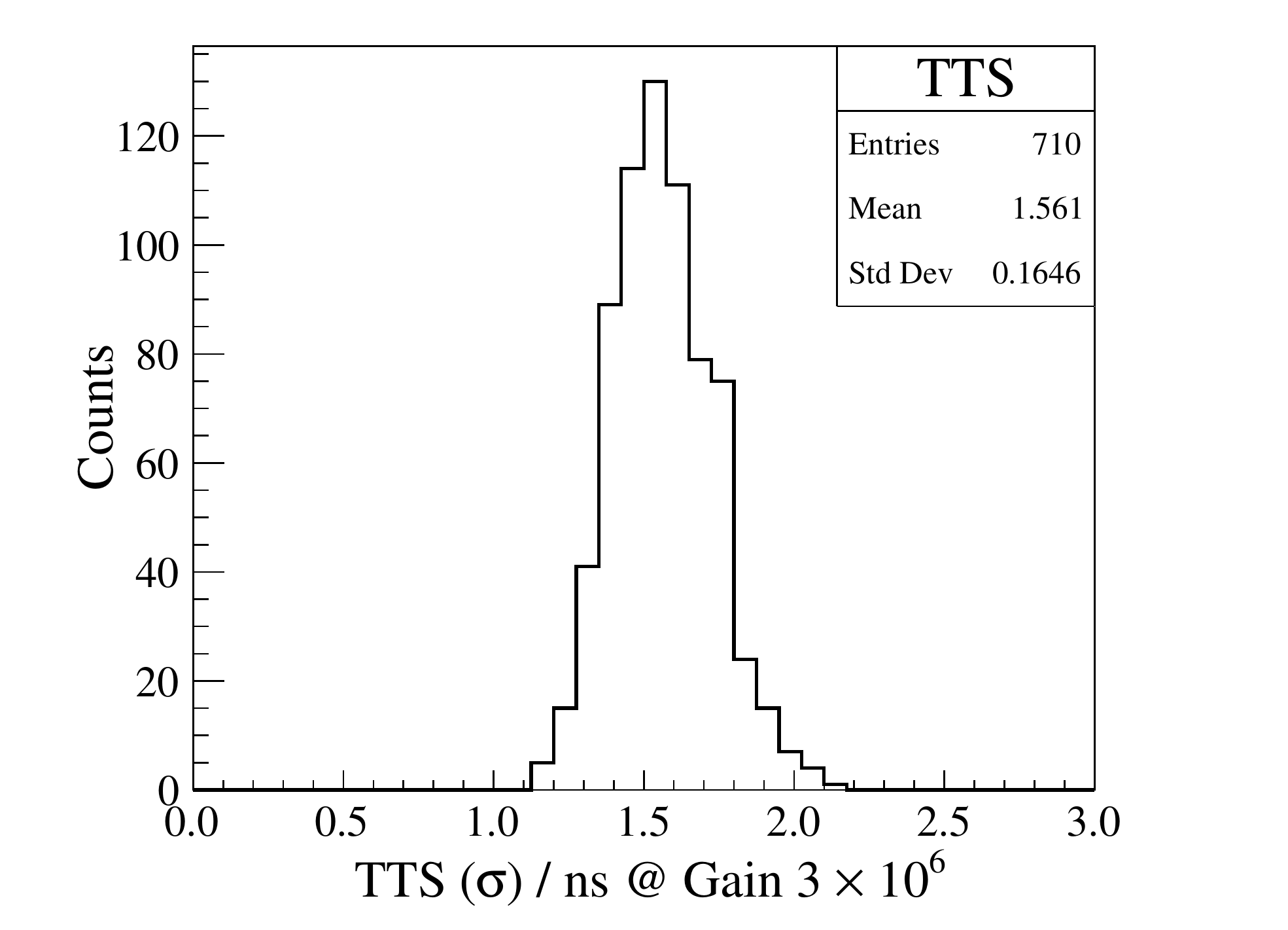}
  \includegraphics[width=0.3\textwidth]{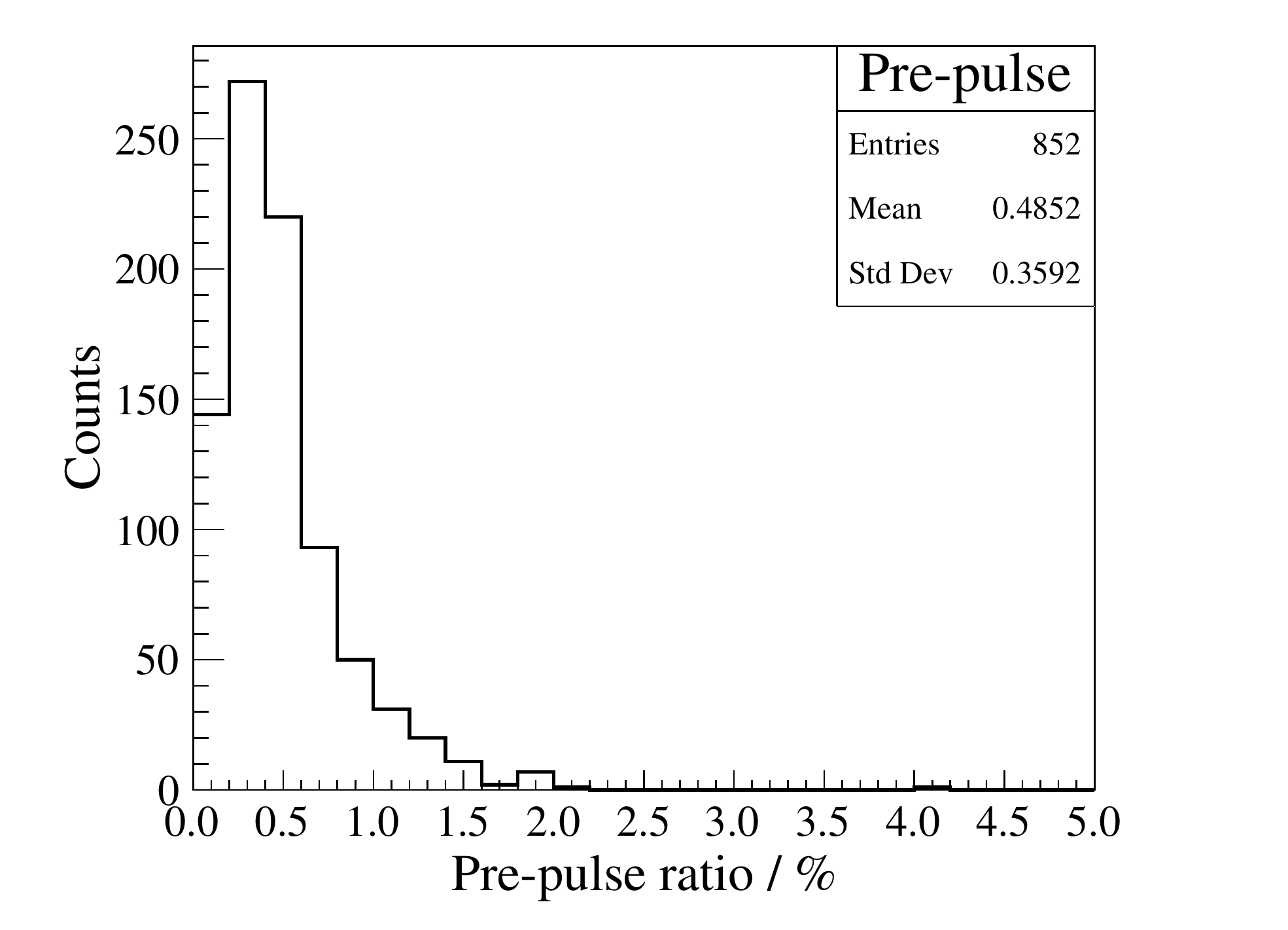}
  \includegraphics[width=0.3\textwidth]{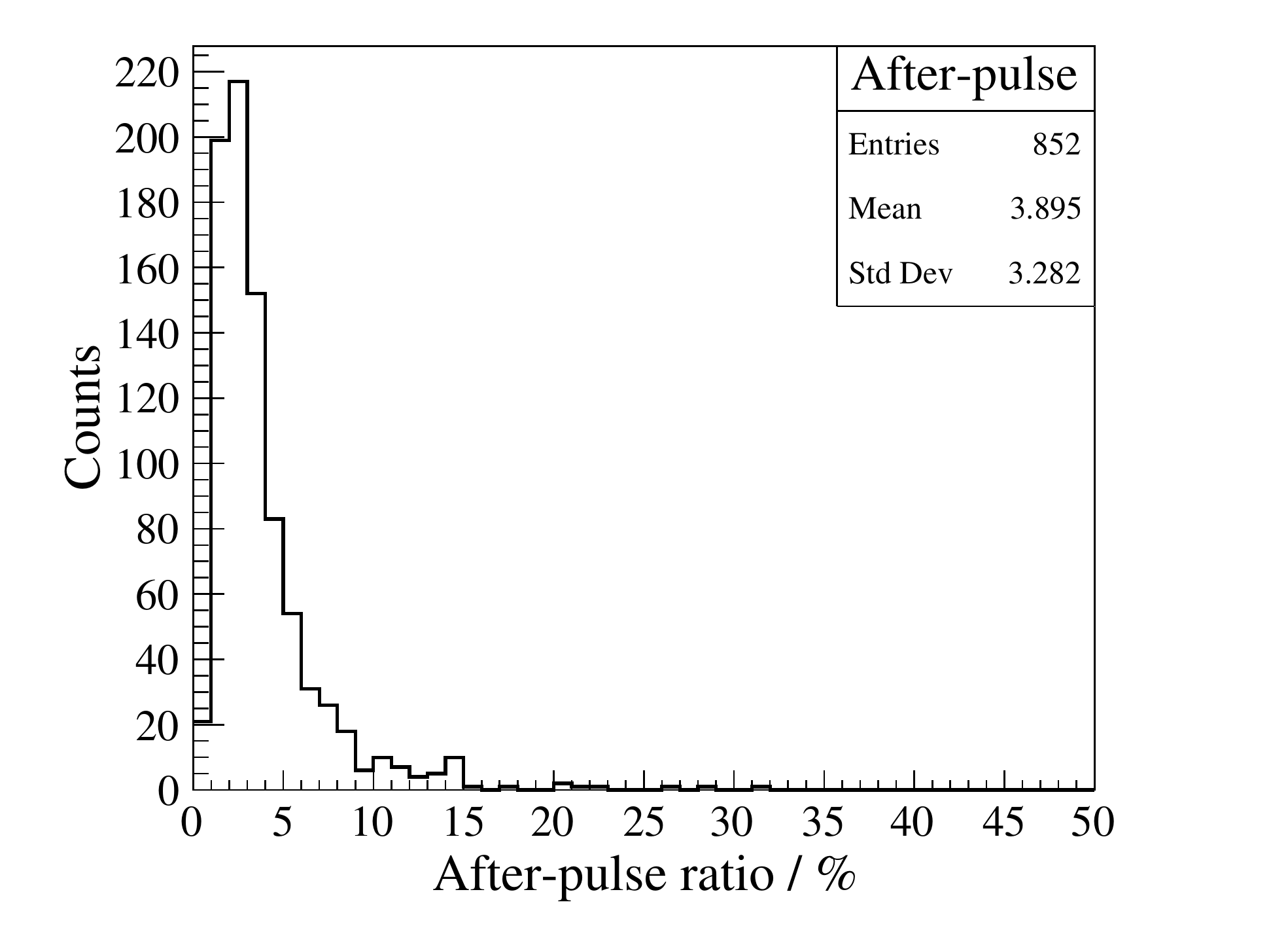}
  \caption{Distribution of TTS, pre-pulse and after-pulse from the sampling test. The number of entries of the TTS plot is less than the other two because the first tens of PMTs were measured at a HV ratio of 3:1:1 and those data were not used. In addition some statistics was added to the pre/after-pulse results due to the resampling after negative test results.}
  \label{fig:tts_allsampling321}
\end{figure}

   In the TTS station, TTS, pre-pulse, and after-pulse were tested, with the results shown in Fig.~\ref{fig:tts_allsampling321}. At the HV ratio 3:2:1. The TTS distribution was very stable. The average was 1.6~ns and the relative deviation was 10.5\%. No PMTs were rejected because of TTS or pre-pulse. However, 11 PMTs were found to be unqualified due to the after-pulse being larger than 15\%, which represented 1.3\% of all of the tested PMTs including those from re-sampling. Considering that the primary function in JUNO of the 3-inch PMTs is single-photon detection, we concluded that this ratio was acceptable.

\begin{figure}[!hbt]
  \centering  
  \includegraphics[width=0.3\textwidth]{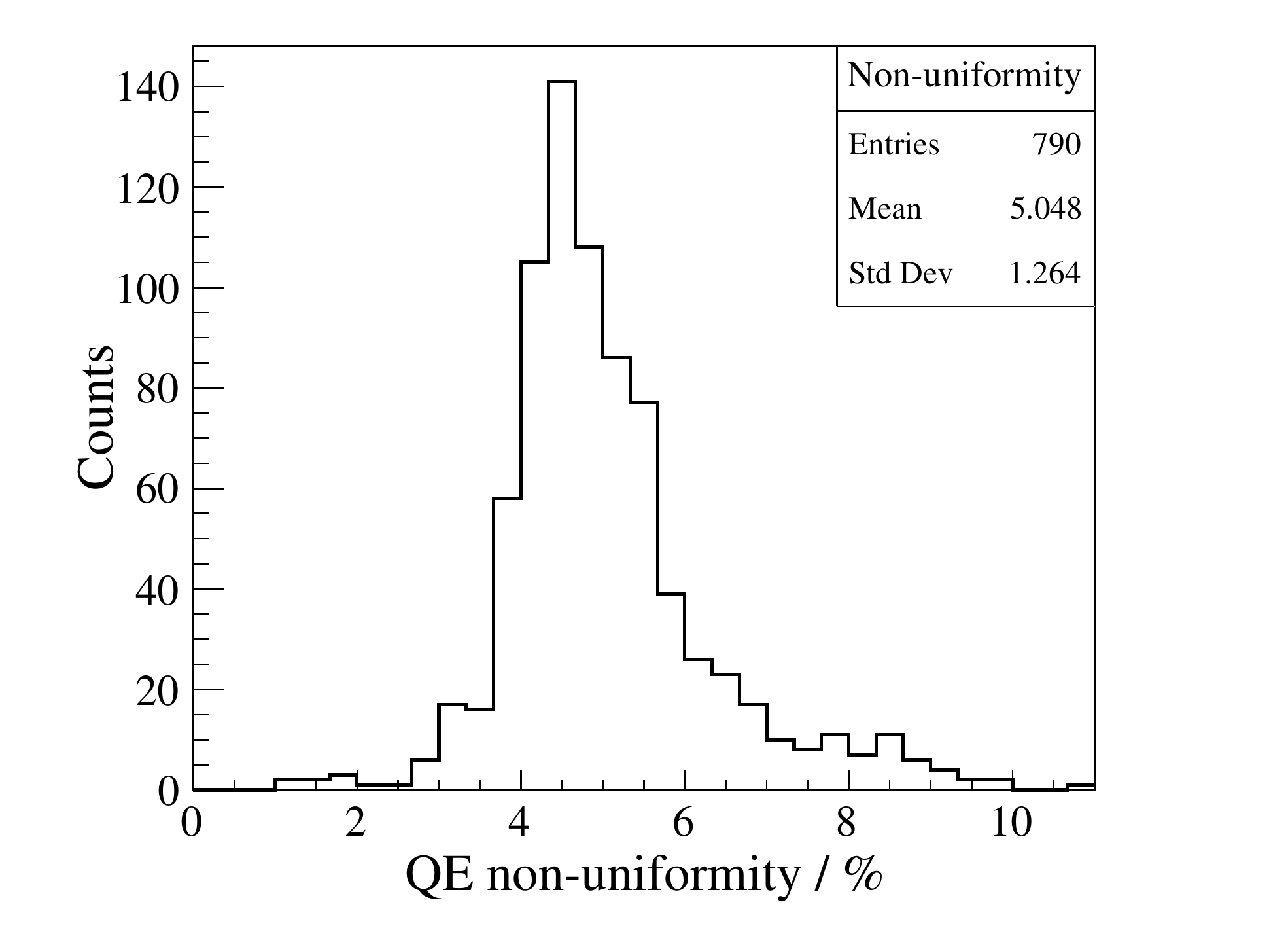}
  \includegraphics[width=0.3\textwidth]{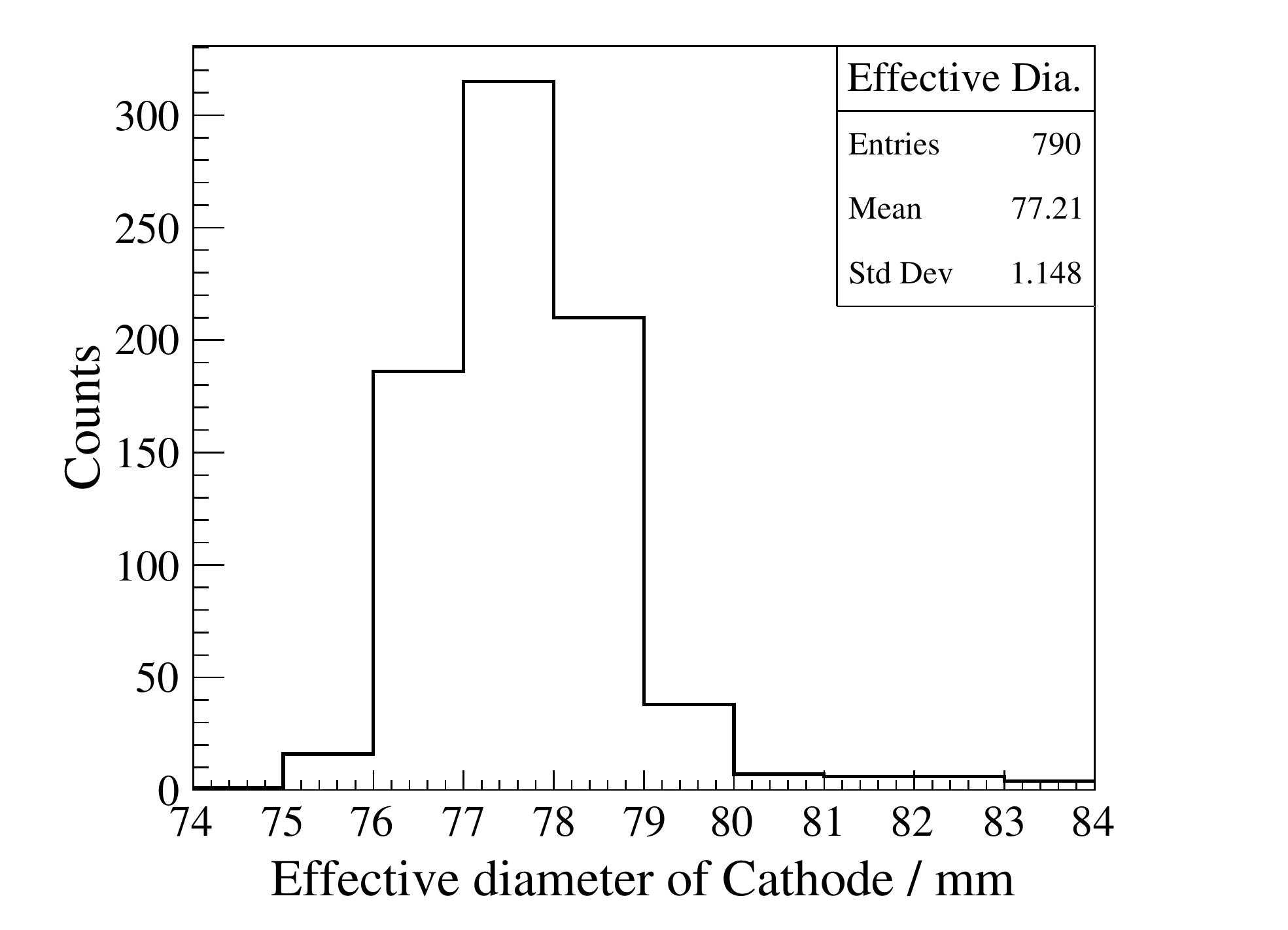}
  \caption{Distribution of QE non-uniformity and effective photocathode diameter from the sampling test. The value in QE non-uniformity plot is the relative percentage of QE by measured the anode current of PMT. Some PMT's effective diameter was larger than 82 mm, because of the uncertainty caused by the 2~mm step length.}
  \label{fig:non-uniform-statistics}
\end{figure}

   In the scanning station, the non-uniformity of the QE and the effective photocathode diameter were evaluated. The results are shown in Fig.~\ref{fig:non-uniform-statistics}. There were 7 PMTs with the QE non-uniformity lower than 2\%. We did an investigation and found they belonged to one batch and were tested in three consecutive days. The 2D scanning map indicated a little light leakage of the test box during that period, which caused a larger mean current value for all pixels and get lower non-uniformity percent value. We concluded they were still qualified. The effective photocathode diameters of a small fraction of PMTs were measured to be larger than 82~mm due to the 2~mm scanning step length. No PMT was rejected at this step.

\begin{figure}[!hbt]
  \centering  
  \includegraphics[width=0.3\textwidth]{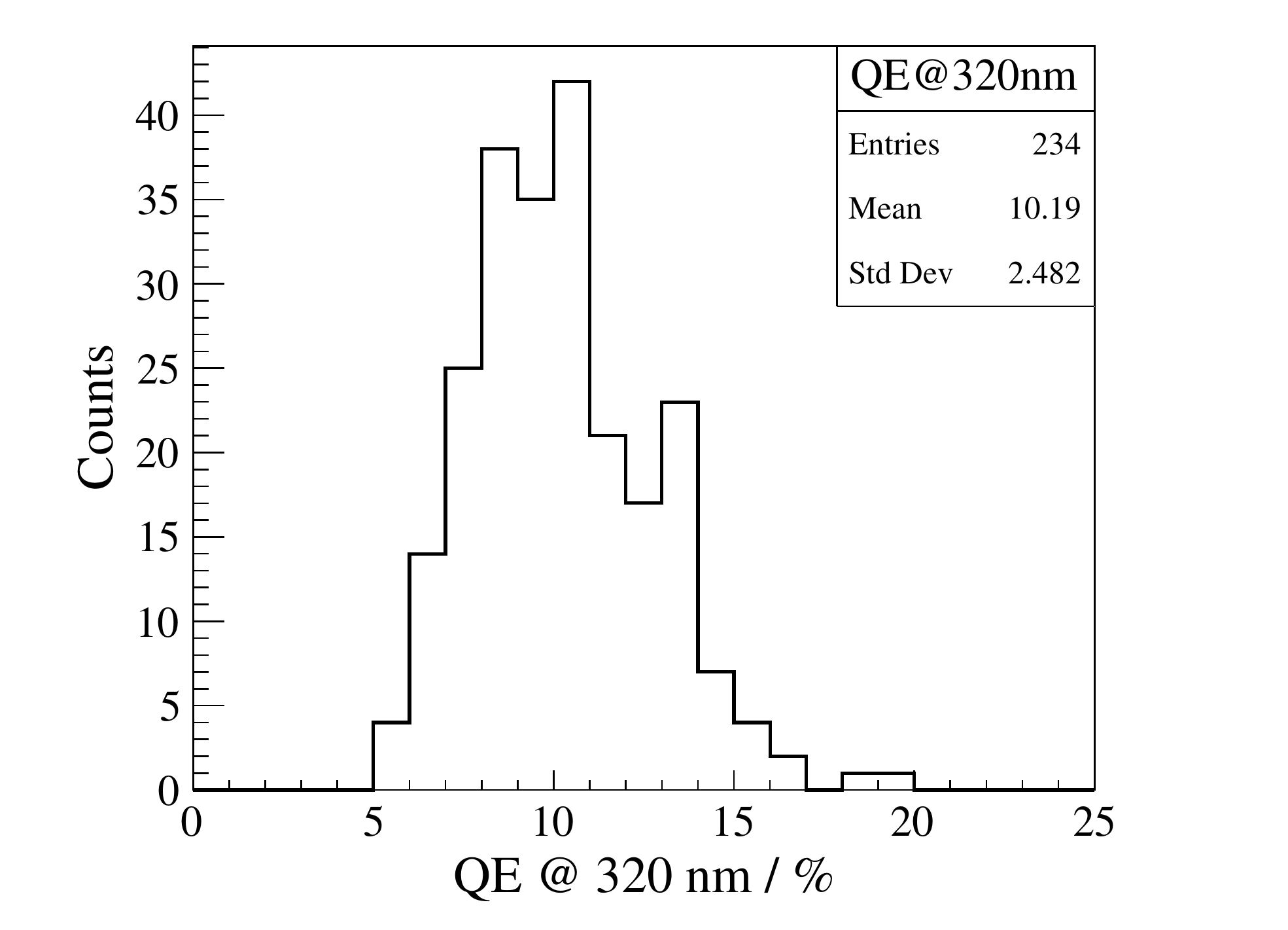}
  \includegraphics[width=0.3\textwidth]{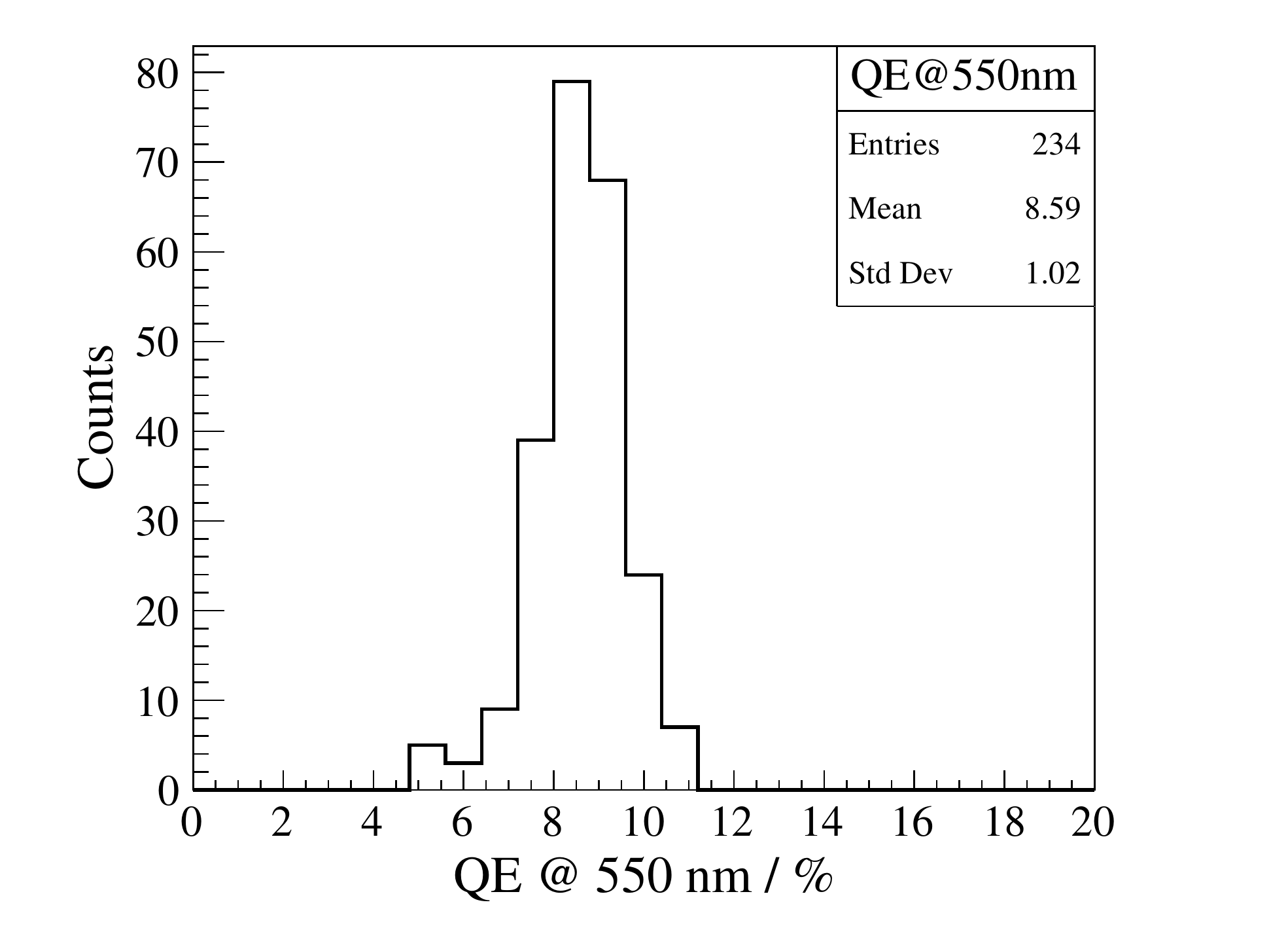}
  \caption{Distribution of QE at light wavelength 320~nm and 550~nm. These two results were used to verify the spectrum response range.}
  \label{fig:qe320-550}
\end{figure}

   To verify the range of the spectrum response, JUNO required the QE at 320~nm and 550~nm larger than 5\%. The measurement was done also in the static station but with different light filters. The results are shown in Fig.~\ref{fig:qe320-550}. All of the sampled PMTs met the requirement.

\subsection{Validation of aging class D parameter in Table 3}

   The PMT gain was expected to decrease as the charge accumulates at the anode. Since JUNO was designed to operate for 20 - 30 years, considering the normal light level in the JUNO detector, the gain decrease was required to be smaller than 50\% with 6.1 coulombs (C) accumulated anode charge, which was calculated from 
   
      \begin{equation}\label{eq:Q}
    Q = R_{\rm{noise}}\times e \times G \times T
   \end{equation}
   Where $Q$ is the charge; $R_{\rm{noise}}$ is the PMT noise, set 2000 Hz here as the maximum noise; $e$ is electron charge, $1.6 \times 10^{-19}$ C; $G$ is the PMT gain, set $3 \times 10^{7}$ as the maximum gain JUNO used in future; $T$ is the time length of PMT working, 20 years.
   
\begin{figure}[!hbt]
  \centering  
  \includegraphics[width=0.5\textwidth]{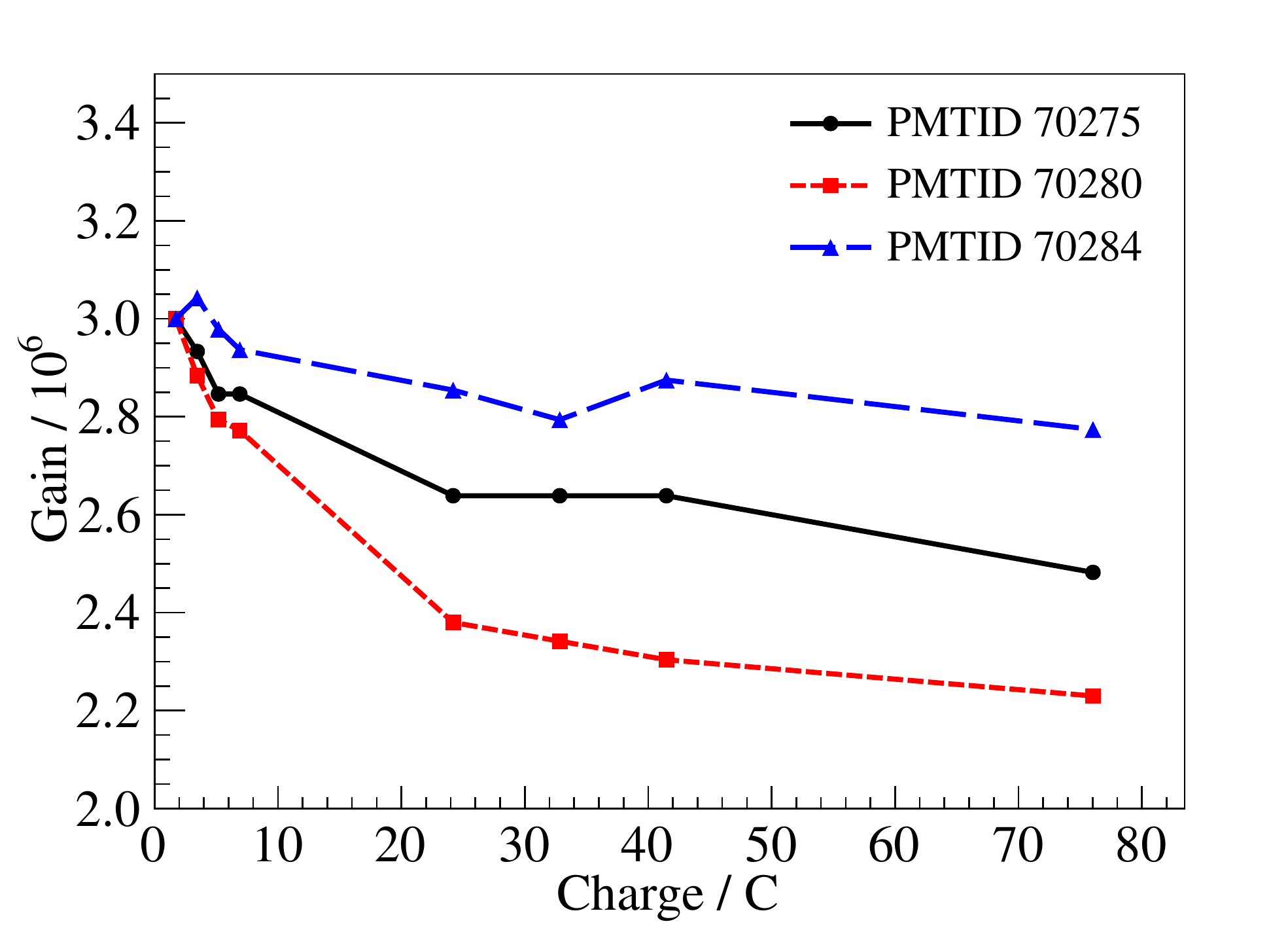}
  \caption{Destructive test to evaluate PMT life time. High intensity light hit the PMT photocathode for several days to simulate the condition of PMTs in JUNO's dark detector with mostly dark noise for 20 years.}
  \label{fig:lifetime}
\end{figure}

   Before mass production, three PMTs were selected for the aging test and exposed to high-intensity light of 10~$\mu$A for 8 days and then 100~$\mu$A for another 8 days continuously, which equals 76 C, about 10 times the JUNO requirement. Their gains were set to $3\times10^6$ in the beginning, and in the end decreased by 8\%, 20\%, and 33\% (Fig.~\ref{fig:lifetime}), respectively, while the QE of each PMT essentially did not change. This meets greatly JUNO requirements.

\subsection{Radioactivity monitoring}
\label{sec:radioactivity}
\begin{figure}[htb!]
		\begin{center}
			\includegraphics[width=0.5\textwidth]{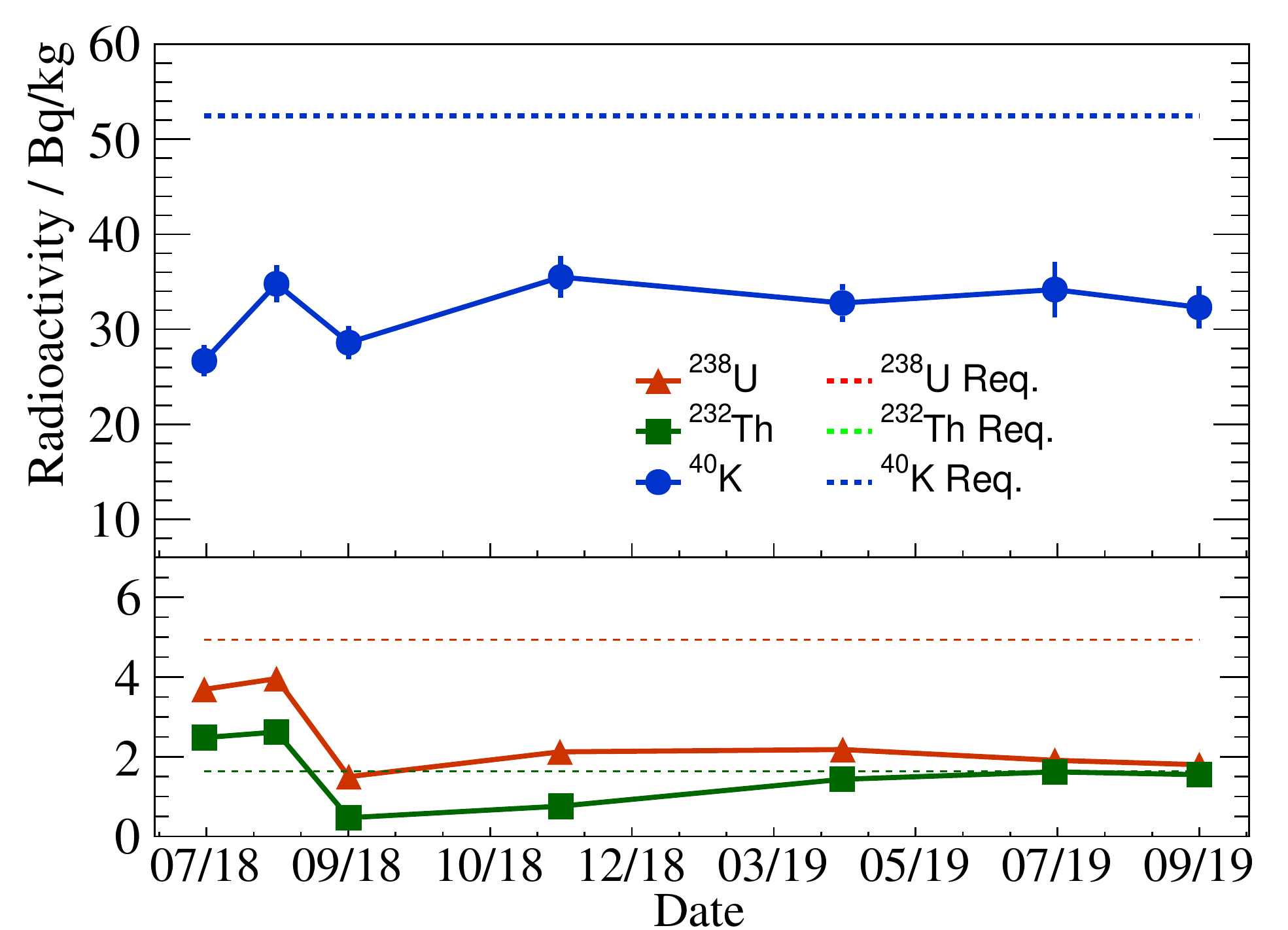}
			\caption{Radioactivity measurements for the glass bulb sample. The dash lines represent the acceptance upper limits.}\label{fig:background}
		\end{center}
	\end{figure}

   During the PMT mass production, the radioactivity of the glass bulb was continuously monitored. The glass bulbs were produced roughly every three months as a batch, and a sample of each batch was sent to JUNO for the radioactivity measurement. There were 7 batches in total and the results are shown in Fig.~\ref{fig:background}. The first two batches were received in the middle of 2018, and $^{232}$Th was found to exceed the acceptance criteria by 50\%-60\%. Considering that the overall background contribution from 3-inch PMTs is very small, these two batches were still accepted. On the other hand, an investigation of the glass bulb factory was done, where the production environment and the procedures were carefully reviewed. In the end, the production was moved to another furnace, and a new stainless steel container was used for the mixing and storage of the raw material (quartz sand, borax, boric acid, aluminum hydroxide, and other minor components) to reduce the dust contamination from the environment. The new sample from the following batch was received one month later and both $^{238}$U and $^{232}$Th were reduced by a factor of 2. After that, later batches showed good stability below the acceptance criteria in Table \ref{tab:rawmaterial} for all of the three elements. 
	
\section{Summary}

   3-inch PMTs will work as an independent photon detection system in JUNO to reduce the systematic uncertainties associated with the energy measurement, improve the energy reconstruction and enhance the physics capabilities. HZC, a Chinese company that has imported the production line from PHOTONIS France, was chosen as the supplier of all 26,000 3-inch PMTs for JUNO. The mass production took 2 years at a speed of $\sim$1,000 pieces per month. The fraction of good PMTs before delivering to JUNO increased continuously and the average yield was 80.5\%. A fraction of the PMT performance parameters were characterized by HZC with two test stations during the production. These parameters were validated by JUNO with a sampling test in the factory. The other parameters were characterized by JUNO with another two test stations also at HZC. In the end, only 15 PMTs were found to be unqualified and thus rejected. 11 of them were rejected due to the after-pulse charge ratio being larger than 15\% and the rest due to having a low working HV, high DCR, and unstable QE. This means that the unqualified PMT ratio was 1.3\% for the after-pulse and 0.15\% for the sum of all other parameters. These results have a negligible impact on the JUNO physics and show the good quality of the PMTs. The radioactivity of the glass bulb was reduced and monitored continuously to meet the requirements of JUNO. All of the measured PMT parameters were stored in the JUNO PMT database~\cite{database}, so that they can be accessed and used by the collaboration during detector installation and commissioning, and eventually for the data analysis. The onsite test by JUNO also provided great help to HZC for better quality control, and the positive experience can be taken as a good reference to other experiments or factories.

\section{Acknowledgments}
   We thank the JUNO low-background working group for radioactivity measurements for the glass bulb. This work was supported by the National Natural Science Foundation of China No. 11975258 and 12005044, the Strategic Priority Research Program of the Chinese Academy of Sciences, Grant No. XDA10011200, the CAS Center for Excellence in Particle Physics, the Funds for Major Science and Technology Programs of Hainan Province (project number: ZDKJ2017011), the Hainan Science and Technology Department, the Special Fund of Science and Technology Innovation Strategy of Guangdong Province, the MOST and MOE in Taiwan, the National Research and Development Agency of Chile, Institut National de Physique Nucl\'eaire et de Physique de Particules (IN2P3) in France, and the University of California at Irvine.



\begin{thebibliography}{99}

\bibitem{juno-CDR}
T. Adam et al., JUNO Conceptual Design Report, arXiv:1508.07166 (2015)
\bibitem{juno-yellowbook}
F. An et al., Neutrino Physics with JUNO, J. Phys. G 43, 030401 (2016)
\bibitem{SPMT-hem}
Miao He, Radiation Detection Technology and Methods, 2017, 1: 21.
\bibitem{km3}
S. Adri\'an-Mart\'inez et al., [KM3Net collaboration], J. Phys. G 43 (2016) 084001
\bibitem{km3-spmt}
S. Aiello et al., [KM3Net collaboration], Journal of Instrumentation. 13. P05035 (2018)
\bibitem{HyperK}
K. Abe et al. (Hyper-Kamiokande Proto-collaboration), arXiv:1805.04163, 2018
\bibitem{HyperK-spmt}
Benjamin Quilain, [Hyper-Kamiokande Proto-collaboration] talk at NEPTUNE conference, 2018
\bibitem{XP72B20}
http://www.hzcphotonics.com/products/XP72B20.pdf
\bibitem{R12199}
https://www.hamamatsu.com/jp/en/product/type/R12199/index.html
\bibitem{linan}
Nan Li et al., Radiation Detection Technology and Methods, 2019, 3(1): 6.
\bibitem{XIOPM}
Guo Le-Hui, Tian Jin-Shou, Lu Yu, and Li Hong-Wei, Acta Phys. Sin. Vol. 65, No. 22 (2016) 228501 (in Chinese)
\bibitem{spmt-base}
Gang Wang, Nuclear Techniques, Vol. 41, No. 8, August 2018 (in Chinese)
\bibitem{pmt-bkg}
Xuantong Zhang, Jie Zhao, Shulin Liu, Shunli Niu, Xiaoming Han, Liangjian Wen, Jincheng He, Tao Hu, NIMA, Volume 898, 1 August 2018, Pages 67-71
\bibitem{radon1}
F. Perrot, Status of SuperNEMO Demonstrator, 38th International Conference on High Energy Physics (ICHEP 2016), PoS (ICHEP2016) 499, https://pos.sissa.it/282/499/pdf
\bibitem{radon2}
C. Cerna, B. Soule and F. Perrot, 
Low Radioactivity Techniques 2015 (LRT2015), AIP, Conference Proceedings 1672, 050002 (2015); http://dx.doi.org/10.1063/1.4927987
\bibitem{PDreference}
http://www.hamamatsu.com/resources/pdf/ssd/s2744-08$\_$etc$\_$kpin1049e06.pdf
\bibitem{QEcalib}
J. Xia et al., 2015 JINST 10 P03023
\bibitem{database}
JUNO PMT database (internal), http://pmtdb.juno.ihep.ac.cn/index.html

  

\end{thebibliography}
\end{document}